\documentclass{aa}  
\usepackage{graphicx}
\usepackage{hyperref}
\usepackage{txfonts}
\usepackage{natbib}
\usepackage{xcolor}
\newif\ifrevision
\revisiontrue
\ifrevision
  \newcommand{\rev}[1]{#1}
\else
  \newcommand{\rev}[1]{#1}
\fi

\newcommand{\kms}{\mbox{km\,s$^{-1}$}}
\newcommand{\hii}{\mbox{H\,{\sc ii}}}
\newcommand{\vlsr}{\mbox{$V_\text{lsr}$}}
\newcommand{\tcsff}{$^{13}\mathrm{CS}$ (5--4)}
\newcommand{\tcs}{$^{13}\mathrm{CS}$}
\newcommand{\ntdptt}{$\mathrm{N_2D^+}$ (3--2)}
\newcommand{\ntdp}{$\mathrm{N_2D^+}$}
\newcommand{\dcntt}{DCN (3--2)}

\makeatletter
\renewcommand{\@biblabel}[1]{}
\makeatother

\makeatletter
\renewcommand*\maketitle{%
  \thispagestyle{firstpage}%
  \begingroup
    \if@wideboxfn
      \setlength\bibindent{1.4\parindent}%
    \else
      \setlength\bibindent{\parindent}%
    \fi
    \renewcommand*\thefootnote{\@fnsymbol\c@footnote}%
    \renewcommand\@makefntext[1]{%
      \ifaa@longfn\hsize\textwidth\fi
      \noindent
      \hb@xt@\bibindent{\hss\@makefnmark\enspace}##1}%
    \ifaa@twocolumn
      \begingroup
        \begin{aa@strip}
          \aa@maketitle
        \end{aa@strip}
        \@thanks
      \endgroup
    \else
      \begingroup
        \let\thanks\footnote
        \aa@maketitle
      \endgroup
    \fi
  \endgroup
  \aa@savethanks \csname c@footnote\endcsname \relax
  \let\@thanks\@empty
  \setcounter{footnote}{0}%
}
\makeatother

\begin{document}

    \title{Dense Cores in the Vicinity of an \hii{} Region}

    \author{Ruofei Zhang\inst{1,2}
            \and 
            Xing Lu\inst{2}
            \and
            Jingwen Wu\inst{1,3}
            \and
            Sihan Jiao\inst{3}
            \and
            Hauyu Baobab Liu\inst{4,5}
            \and
            Guang-Xing Li\inst{6}
            \and
            Roberto Galván-Madrid\inst{7}
            \and
            Aiyuan Yang\inst{3,8}
            \and
            Siju Zhang\inst{9,10}
            \and
            Shanghuo Li\inst{11}
            \and
            Fengwei Xu\inst{12}
            \and
            Xindi Tang\inst{13}
            \and
            Yu Cheng\inst{14}
            \and
            Weiyuan Zhang\inst{15}
            \and
            Andrés E. Guzmán\inst{16}
            \and 
            Yuxin Lin\inst{17}
            \and
            Yuhua Liu\inst{2}
            \and
            Qizhou Zhang\inst{18}
            \and
            Patricio Sanhueza\inst{19}
            \and
            Ke Wang\inst{20}
            \and
            Siyi Feng\inst{21}
            \and
            Linjing Feng\inst{1,3}
            \and
            Fangyuan Deng\inst{1,3}
            \and
            Hao Ruan\inst{1,3}
            \and
            Yuanzhen Xiong\inst{1,3}
            \and
            Yuxiang Liu\inst{1,3}
            }

    \institute{
            University of Chinese Academy of Sciences, Beijing 100049, China
        \and
            Shanghai Astronomical Observatory, Chinese Academy of Sciences, 80 Nandan Road, Shanghai 200030, PR China
            \rev{\email{xinglu@shao.ac.cn}}
        \and
            National Astronomical Observatories, Chinese Academy of Sciences, Beijing 100101, People's Republic of China
            \rev{\email{jingwen@nao.cas.cn}}
        \and
            Physics Department, National Sun Yat-Sen University, Kaohsiung City 80424, Taiwan
        \and
            Center of Astronomy and Gravitation, National Taiwan Normal University, Taipei 116, Taiwan
        \and
            South-Western Institute for Astronomy Research, Yunnan University, Chenggong District, Kunming 650091, China
        \and
            Instituto de Radioastronomía y Astrofísica, Universidad Nacional Autónoma de México, Morelia, Michoacán 58089, Mexico
        \and
            Key Laboratory of Radio Astronomy and Technology, Chinese Academy of Sciences, A20 Datun Road, Chaoyang District, Beijing, 100101, People's Republic of China
        \and
            Departamento de Astronomía, Universidad de Chile, Camino el Observatorio 1515, Las Condes, Santiago
        \and
            Chinese Academy of Sciences South America Center for Astronomy, National Astronomical Observatories, Chinese Academy of Sciences, Beijing 100101, People’s Republic of China
        \and
            School of Astronomy and Space Science, Nanjing University, Nanjing 210093, People’s Republic of China
        \and
            Max Planck Institute for Astronomy, Konigstuhl 17, D-69117 Heidelberg, Germany
        \and 
            XingJiang Astronomical Observatory, Chinese Academy of Sciences(CAS), Urumqi 830011, PR China
        \and
            National Astronomical Observatory of Japan, 2-21-1 Osawa, Mitaka, Tokyo 181-8588, Japan
        \and
            Shanghai Pinghe School, 333 Shenqi Road, Pudong District, Shanghai, China
        \and
            Joint Alma Observatory (JAO), Alonso de Córdova 3107, Vitacura, Santiago
        \and
            Max-Planck-Institut für Extraterrestrische Physik, Giessenbachstr. 1, 85748 Garching bei München, Germany
        \and
            Center for Astrophysics | Harvard \& Smithsonian, 60 Garden Street, Cambridge, MA 02138, USA
        \and
            Department of Astronomy, School of Science, The University of Tokyo, 7-3-1 Hongo, Bunkyo, Tokyo 113-0033, Japan
        \and
            Kavli Institute for Astronomy and Astrophysics, Peking University, Beijing 100871, China
        \and
            Department of Astronomy, Xiamen University, Zengcuo’an West Road, Xiamen, 361005, Peoples Republic of China
        }

    \abstract
    {Massive stars form in the densest part of molecular clouds and strongly influence their surroundings through radiative and mechanical feedback. The effects of such feedback on dense gas structures at sub-pc scales, however, remain to be better constrained by observations.}
   {We aim to investigate how feedback from a newly formed massive star affects the physical properties of dense cores embedded in a neighboring filamentary molecular cloud.}
   {We analyze ALMA Band~6 observations of the filamentary source IRAS~18530+0215 (I18530), including 1.3 mm dust continuum, molecular line emission from DCN, \ntdp{}, and \tcs{}, as well as complementary VLA K-band continuum and NH$_3$ meta-stable line observations. The dynamical state of the ultra-compact (UC)~\hii{} region is investigated through energy and pressure estimates. Dense cores traced by the line emission are identified using the \textit{astrodendro} algorithm, and their temperatures, masses, velocity dispersions, and virial parameters are derived from molecular line fitting and continuum emission. We further examine how the physical and kinematic properties of the dense cores vary under the influence of feedback from the \hii{} region.}
   {The \hii{} region has a radius of $\sim$0.1~pc and exhibits an expansion velocity of $\sim$2.5~\kms{}, corresponding to a shell dynamical age of $\sim$0.06~Myr. DCN and \tcs{} cores are primarily distributed in the vicinity of the \hii{} region, whereas \ntdp{} cores are preferentially found at larger distances. Overall, the temperatures and velocity dispersions of the dense cores show clear decreasing trends with increasing projected distance from the \hii{} region. The virial parameters increase with projected distance within the inner $\sim0.3$~pc, but decline sharply beyond this scale, while the core masses show no significant variation with distance. Strong star formation signatures are found at projected distances of $\sim$0.2~pc from the \hii{} region, whereas more distant regions still host quiescent, cold dense cores at earlier evolutionary stages.}
   {Our results suggest that the compact \hii{} region is currently trapped or choked within a scale of $\sim0.1$~pc. Its feedback extends to at least $\sim0.3$~pc and is unlikely to significantly disrupt the entire filamentary structure on short timescales. Within the affected region, the feedback enhances the velocity dispersions, gas temperatures, and virial parameters of the dense cores, with no evidence that it promotes the formation of more massive dense cores.}


    \maketitle
\nolinenumbers

\section{Introduction}

    Observations of molecular clouds in the Galaxy have established that massive stars ($M>8~M_{\odot}$) are preferentially formed in the densest part of molecular clouds, which are often associated with filamentary structures and hubs \citep[e.g.,][]{Schneider2012,Liu2012_1,Lin2016,NandaKumar2020,Hacar2023,Pineda2023,GXLi2025}.
    The formation of massive stars and their subsequent feedback constitute one of the central problems in contemporary star formation studies \citep{Zinnecker2007,Beuther2025}. Unlike low-mass stars, massive stars release substantial amounts of energy already at very early evolutionary stages through intense ultraviolet radiation, stellar winds, and bipolar outflows. These feedback processes can profoundly reshape their natal molecular clouds, altering gas density distributions, velocity fields, and the evolutionary pathways of subsequent star formation \citep{Krumholz2014}.

    The impact of massive star feedback on surrounding molecular gas is inherently complex. On the one hand, feedback-driven compression can promote secondary collapse and trigger new generations of star formation \citep[e.g.,][]{Elmegreen1998,Luisi2021,Dale2025}. On the other hand, feedback may disperse dense gas, accelerate cloud dispersal, and suppress star formation by removing the material required for further collapse \citep[e.g.,][]{Bending2020,Lewis2023}.
    Both positive and negative feedback signatures have been reported in observations and numerical simulations, indicating that the outcome of feedback is strongly environment-dependent. In particular, small-scale accretion structures may help sustain accretion by reducing the impact of radiative feedback at early evolutionary stages \citep[e.g.,][]{Peters2010,Kuiper2018}. The efficiency and mode of feedback are regulated by multiple factors, including the initial cloud structure, density distribution, turbulent state, and evolutionary stage \citep[e.g.,][]{Matzner2015,Dale2017,Sartorio2021,Fukushima2021}.
    
    While numerous observational studies have demonstrated that massive star feedback can significantly influence the structure and evolution of molecular clouds, most of these works have primarily focused on relatively large spatial scales ($\gtrsim$ a few~pc) or more evolved regions \citep[e.g.,][]{McLeod2019,Luisi2021,Xu2026}. As a result, the detailed impact of feedback on the internal substructures of molecular clouds, such as dense filaments, cores, and local velocity fields, remains poorly constrained. Recent high-resolution observations with ALMA have begun to probe feedback processes down to sub-parsec scales, revealing that massive star feedback can reshape gas distributions, alter fragmentation behavior, and introduce complex kinematic signatures \citep[e.g.,][]{Zhang2020,Motte2022,Zhang2023,Zhang2024,Liu2024,Armante2024,Guo2025,Zhen2025}. Despite these advances, it is still unclear how feedback couples to dense gas at these small scales, and to what extent it regulates the physical and kinematic properties of individual cores and filamentary structures. Investigating feedback at sub-pc scales is therefore essential for understanding how star formation is regulated in clustered environments.

    \begin{figure}[ht]
        \centering
        \includegraphics[width=0.5\textwidth]{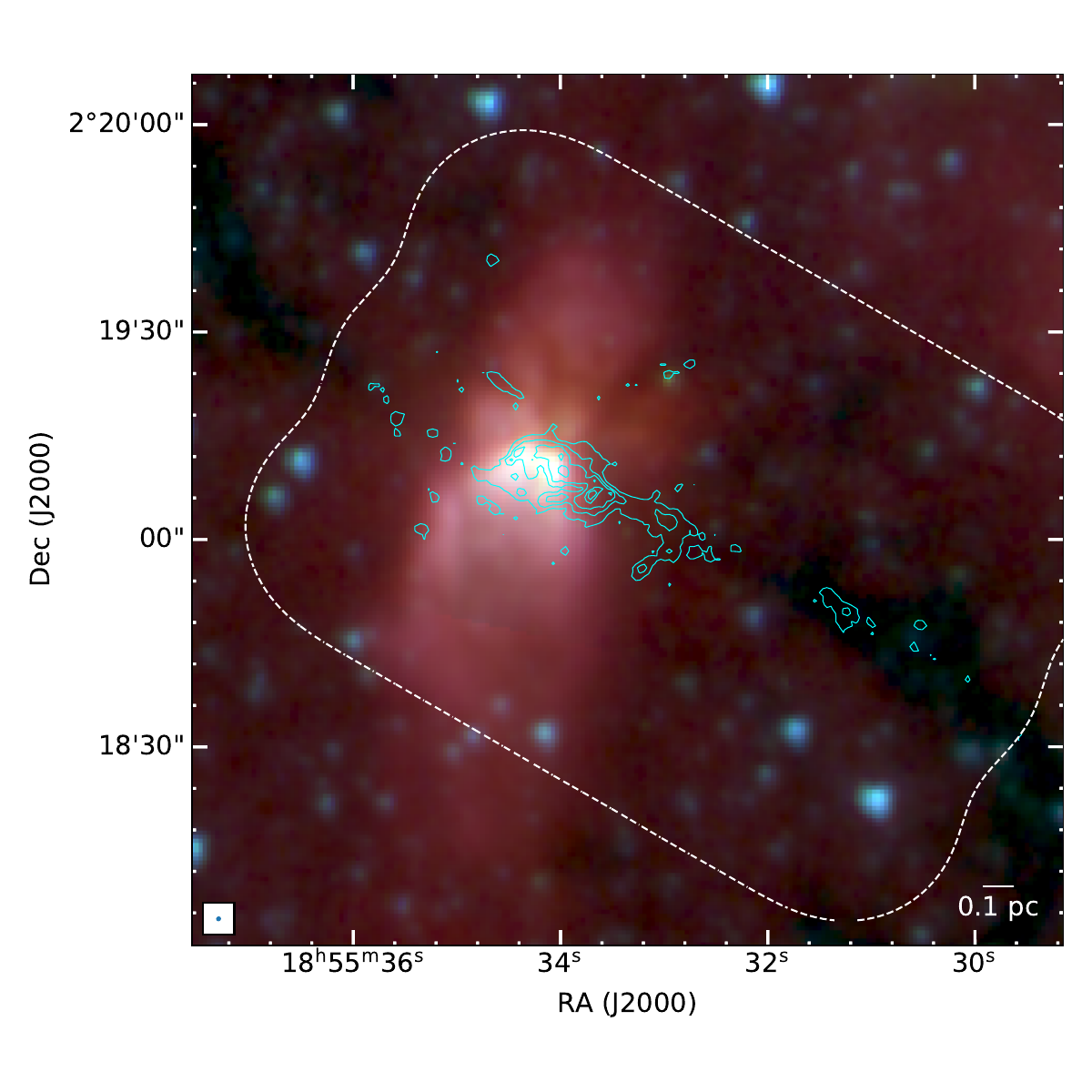}
        \caption{Three-color composite image of the I18530 cloud. The red, green, and blue channels represent the Spitzer/IRAC 8.0~$\mu$m, 4.5~$\mu$m, and 3.6~$\mu$m emission, respectively. Cyan contours indicate the ALMA 1.3 mm continuum emission at 6, 20, 40, 100, and 160 times the RMS noise level of 0.11~Jy~beam$^{-1}$. The white dashed contour delineates the field covered by the ALMA 1.3 mm mosaic observations. The synthesized beam ($0\farcs75 \times 0\farcs70$) is shown in the lower left corner.
}
        \label{fig:RGB}
    \end{figure}

    In this context, the source IRAS~18530+0215 (hereafter I18530), observed as part of the Investigations of Massive Filaments and Star Formation (INFANT) project \citep{Cheng2024}, provides an excellent laboratory for studying massive star feedback at sub-pc scales. I18530 is a filamentary molecular cloud at a distance of 4.6~kpc \citep{Cheng2024,Lu2018}, exhibiting a clear evolutionary gradient: an ultra-compact (UC)~\hii{} region, powered by a newly formed massive star, is located near the central part of the filament, while both ends remain infrared dark (\autoref{fig:RGB}), indicating earlier evolutionary stages.
    This configuration allows the feedback from the newly formed massive star to be studied within a relatively simple and well-defined geometry, where dense gas structures coexist with an ionized region on scales below 1~pc. The availability of high-resolution ALMA Band~6 observations, combined with VLA K-band data, enables us to resolve individual dense cores down to a spatial resolution of approximately 3000~AU, making it possible to directly assess how massive star feedback influences nearby dense gas.

    The goal of this work is to investigate the impact of massive star feedback on dense gas structures within the filamentary cloud I18530 at sub-pc scales. By combining ALMA molecular line data tracing gas at different temperatures and densities with VLA continuum observations of ionized gas, we aim to characterize the physical properties and spatial distribution of dense cores and to assess how feedback from the UC~\hii{} region affects their thermal, kinematic, and dynamical states.
    This paper is organized as follows: Section~\ref{sec:obs} presents the observational data. Section~\ref{sec:results} describes the analysis and results. Section~\ref{sec:disc} analyzes the spatial distribution of dense gas traced by different molecular and continuum diagnostics and discusses the impact of massive star feedback on the surrounding environment. Finally, Section~\ref{sec:conc} summarizes our major findings. 

\section{Observation and Data Reduction}\label{sec:obs}

    \begin{table*}
        \caption{Summary of observational data used in this work.}
        \label{tab:obs_summary}
        \centering
        \begin{tabular}{lccccc}
        \hline\hline
        Telescope & Band / Line & Velocity Resolution & Beam Size & Sensitivity (mJy beam$^{-1}$) \\
        \hline
        ALMA & Band 6 continuum & -- & $0\farcs75 \times 0\farcs70$ & 0.11 (4.7~mK) \\
        ALMA & \dcntt, H$_2$CO, \tcsff{}, \ntdptt{} 
        & 0.63--0.67~\kms{} & $0\farcs85 \times 0\farcs80$ 
        & 4.8--6.8 (0.17--0.24~K) \\
        \hline
        VLA & K-band continuum & -- & $1\farcs25 \times 1\farcs06$ & 0.015 (0.029~K) \\
        VLA & NH$_3$ (1,1), (2,2) & 0.4~\kms{} & $\sim 1\farcs2$ & 1.3--1.5 ($\sim$2~K) \\
            VLA & H$_2$O maser & 0.4~\kms{} & $\sim 1\farcs2$ & 1.3--1.5 ($\sim$2~K) \\
        \hline
        CSO & 1.1~mm continuum & -- & $33\farcs0$ & -- \\
        SMA & 1.3~mm continuum & -- & $3\farcs4 \times 2\farcs6$ & 1.0 (0.26~mK) \\
        MaserDB & CH$_3$OH maser (Class I) & -- & $1\farcs8 \times 1\farcs6$ & -- \\
        Herschel & 70~$\mu$m, 160~$\mu$m & -- & $5\farcs2$, $12\farcs0$ & -- \\
        \hline
        \end{tabular}
    \end{table*}

    The source I18530 was observed as part of the INFANT project \citep{Cheng2024} using the Atacama Large Millimeter/submillimeter Array (ALMA) in Band~6 during Cycle~5 under project 2017.1.00526.S (PI: X.~Lu). The observations were carried out with the 12~m array in two configurations (C43-4 and C43-1), providing sensitivity to both compact and moderately extended structures. The main observational parameters are summarized in Table~\ref{tab:obs_summary}.

    In addition to the continuum emission, multiple molecular transitions were observed simultaneously in Band~6, including SiO (5--4), CO (2--1), \dcntt, $\mathrm{H_2CO}$ transitions, \tcsff{}, and \ntdptt{}. In this work, we focus on dense gas tracers (\dcntt, \tcsff{}, and \ntdptt{}) to identify dense cores, while the $\mathrm{H_2CO}$ transitions $3_{0,3}$--$2_{0,2}$, $3_{2,2}$--$2_{2,1}$, and $3_{2,1}$--$2_{2,0}$ are used to derive the gas temperature.

    The data were calibrated using the standard pipeline within \texttt{CASA} version 6.6.1 \citep{casa2022}. Imaging was performed with the \texttt{tclean} task using a Briggs weighting scheme with a robust parameter of 0.5. A cell size of 0\farcs{15} was adopted, and multiscale deconvolution was applied with scales of [0, 10, 25, 50] pixels. The same imaging parameters were used for all spectral line cubes to ensure consistency across different tracers.

    We also used K-band data from the Karl G.~Jansky Very Large Array (VLA) presented by \citet{Cheng2024}, including continuum emission and NH$_3$ inversion transitions. The 1.3~cm continuum emission traces free-free emission from the \hii{} region, while the $\mathrm{NH_3}$ (1,1) and (2,2) inversion transitions were used to derive the gas temperature and kinematic properties of the dense gas. The associated $\mathrm{H_2O}$ masers were also included as tracers of ongoing star formation activity.

    To recover the extended dust emission filtered out by the interferometric observations, we combined the ALMA 1.3~mm continuum image with archival Submillimeter Array (SMA) 1.3~mm observations from \citet{Lu2018} and Caltech Submillimeter Observatory (CSO)/Bolocam Galactic Plane Survey (BGPS) 1.1~mm continuum data \citep{Ginsburg2013} retrieved from the IRSA archive (\href{https://doi.org/10.26131/IRSA482}{DOI: 10.26131/IRSA482}) using the feathering technique. The resulting continuum image better recovers the diffuse large-scale structures and was therefore used to estimate the physical properties of the extended structures associated with the \hii{} region, including the shell mass and gas number density. In contrast, the compact dense cores were analyzed using the original high-resolution ALMA continuum image alone.

    In addition, Class~I $\mathrm{CH_3OH}$ masers from the MaserDB database \citep{Ladeyschikov2019} were included as complementary tracers of ongoing star formation activity. Herschel PACS 70 and 160~$\mu$m data were used to estimate the dust-reprocessed radiation pressure \citep{Pilbratt2010,Poglitsch2010}.

\section{Results and Analysis}\label{sec:results}

    \subsection{The filamentary structure revealed by NH$_3$ emission}\label{subsec:filament}
    
\begin{figure}[ht]
        \centering
        \includegraphics[width=0.5\textwidth]{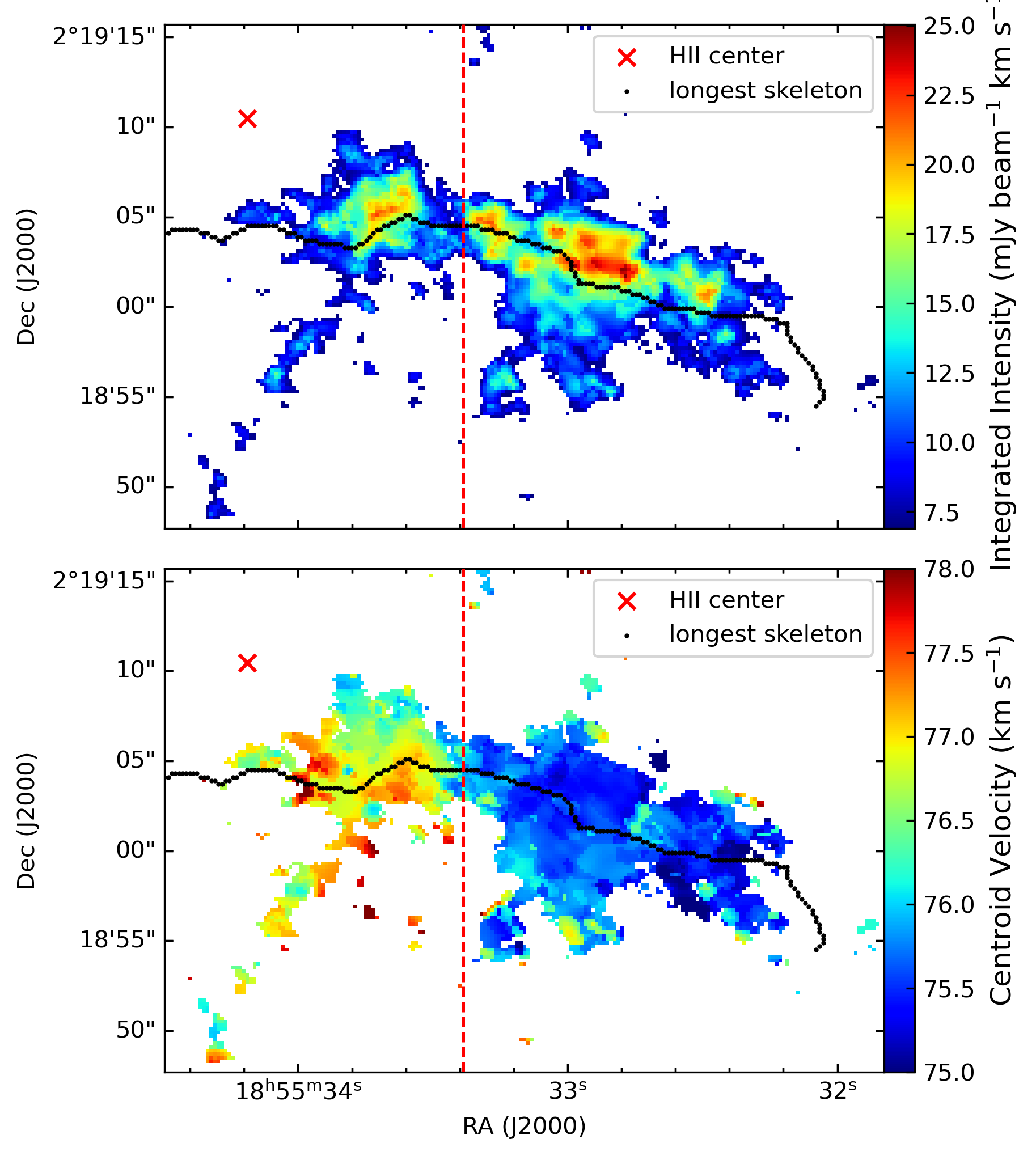}
        \caption{The integrated intensity (top; integrated from 73 to 81~km~s$^{-1}$) and centroid velocity (bottom) map of the NH$_3$ (1,1) main hyperfine component. The black curve indicates the filament skeleton identified by \textit{FilFinder}, while the red cross marks the position of the \hii{} region, derived from a 2D Gaussian fit to the VLA K-band continuum emission. The vertical red dashed line denotes the location corresponding to the velocity discontinuity seen in the PV diagram.}
        \label{fig:NH3 moment}
    \end{figure}

    To investigate the filamentary structure and kinematics in I18530, we first constructed the integrated intensity and centroid velocity maps of the VLA $\mathrm{NH_3}$ (1,1) line (\autoref{fig:NH3 moment}), masking out areas with intensity levels below 3 times the RMS noise, where the RMS is 2.3~mJy~beam$^{-1}$~km~s$^{-1}$. We chose $\mathrm{NH_3}$ as the tracer of the filamentary structure, because its critical density \citep[$\sim$10$^3$~cm$^{-3}$ at a temperature of 20~K;][]{Shirley2015} is below the mean density in the filament \citep[$\sim$10$^4$~cm$^{-3}$, which was estimated from the dust emission;][]{Lu2018} and therefore reveals the filamentary morphology more clearly than other available molecular tracers in our ALMA observations.
    
    We then used the FilFinder package\footnote{\url{https://fil-finder.readthedocs.io}} \citep{Koch2015} to identify filamentary structures in the integrated intensity map. First, we smoothed the image and removed the background by setting all pixels below the 85th percentile of the intensity distribution to a constant background level. Then, we created a mask that excludes the image boundaries and small structures with a size threshold of 80 pixels$^2$ (corresponding to approximately two synthesized beams). Finally, we applied the skeletonization to detect the main filamentary structure. These parameters were chosen to suppress diffuse background emission and isolated small-scale noise-like structures while preserving the dominant filamentary morphology. The longest filament, which is highlighted in \autoref{fig:NH3 moment}, was selected for further analysis.

\begin{figure}[ht]
        \centering
        \includegraphics[width=0.5\textwidth]{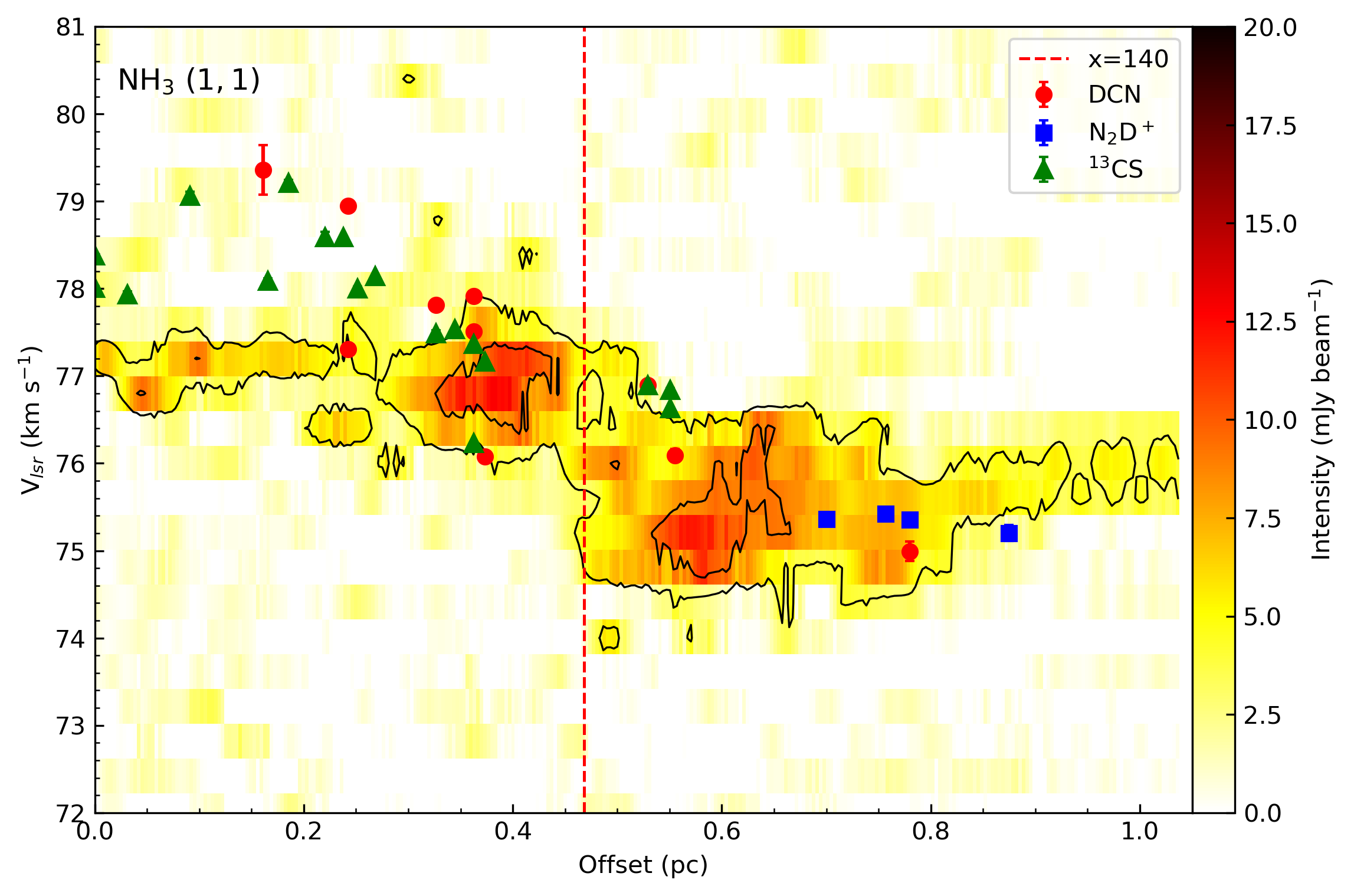}
        \caption{PV diagrams of NH$_3$ (1,1) main line along the filament skeleton. The vertical dashed line marks the position of the velocity discontinuity. Red circles, blue squares, and green triangles denote the centroid velocities of dense cores identified in DCN, \ntdp{}, and \tcs{}, respectively. The contours are plotted at levels of (4, 9, 13)~mJy~beam$^{-1}$}.
        \label{fig:PV diagrams}
    \end{figure}

    We generated position-velocity (PV) diagrams for the $\mathrm{NH_3}$ (1,1) main line along the longest filament using Python package \textit{pvextractor}\footnote{\url{http://pvextractor.readthedocs.io}} (\autoref{fig:PV diagrams}). The PV diagrams reveal a clear discontinuity in \vlsr{} at an offset of 0.47~pc, indicated by a red dashed line in both panels. A similar velocity discontinuity is also detected in the $\mathrm{H_2CO}$ $3_{0,3}$--$2_{0,2}$ transition at the same position. This feature likely marks a transition between two physical components: the left side corresponds to the \hii{} region, while the right side traces a filamentary gas structure. Given the small difference  in \vlsr{} across this discontinuity ($\sim$1.5~\kms{}), two interpretations are possible: the two components may be spatially separated, or the discontinuity may result from localized gas perturbation or dispersal driven by feedback from the \hii{} region. The latter interpretation is favored owing to the small velocity difference and the presence of nearby star formation activity (see \autoref{subsec:OutlierCores}).

    \subsection{ALMA Continuum and Spectral Lines Emission}
    
\begin{figure*}[ht]
        \centering
        \includegraphics[width=1\textwidth]{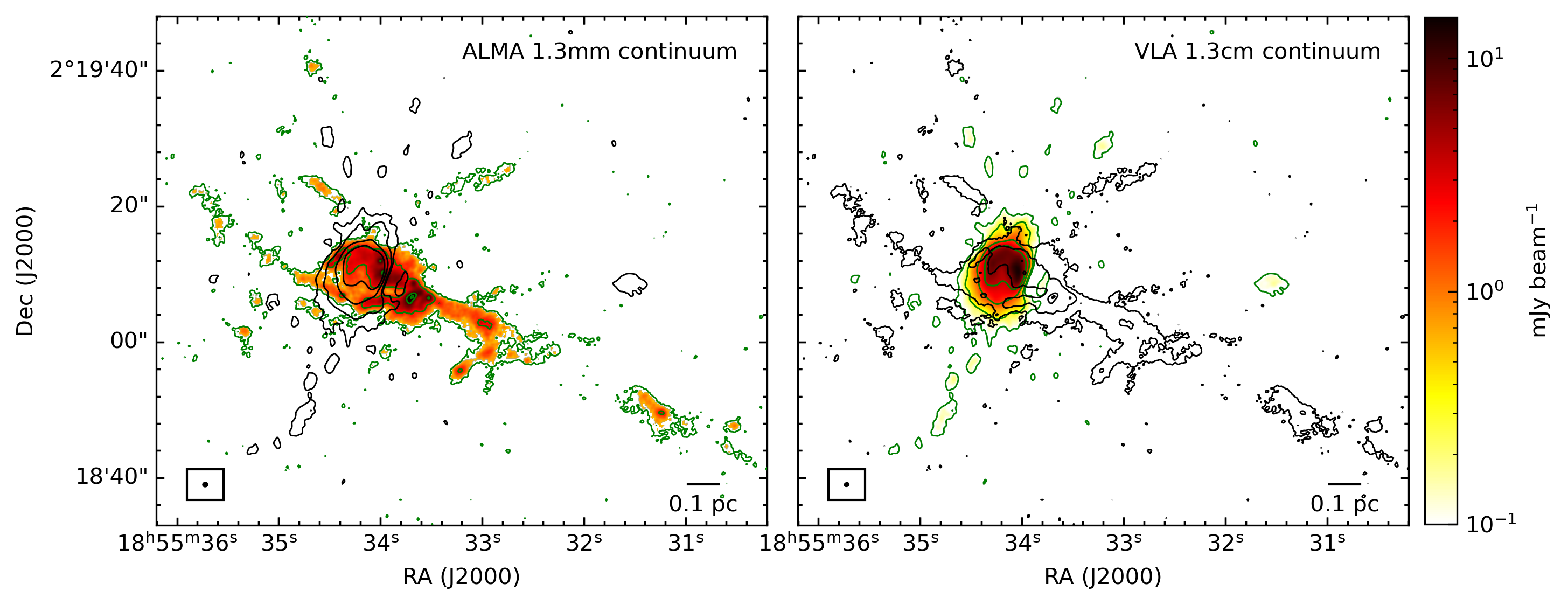}
        \caption{ALMA 1.3~mm~and VLA 1.3~cm continuum emission maps, neither of which is corrected for the primary beam response so as to present a uniform noise level across the field. The green contours show [3, 20, 100, 160]$\times$RMS for the ALMA map (RMS = 0.11~Jy~beam$^{-1}$) and [5, 20, 100, 250]$\times$RMS for the VLA map (RMS = 0.015~Jy~beam$^{-1}$). The black contours correspond to the emission shown in the other panel, overlaid for comparison.}
        \label{fig:continuum}
    \end{figure*}

    \autoref{fig:continuum} shows the ALMA 1.3~mm continuum and VLA 1.3 cm~continuum images of I18530. The 1.3~mm continuum emission is contributed by thermal dust emission and free-free emission. The source exhibits a filamentary structure extending from north-east to south-west. The 1.3~cm continuum emission is mainly attributed to free–free emission from the \hii{} region located in the north-eastern part of the source.

    To further characterize the physical and chemical properties of the dense gas, we make use of several molecular tracers covered by the ALMA observations. The $\mathrm{H_2CO}$ transitions $3_{0,3}$--$2_{0,2}$, $3_{2,2}$--$2_{2,1}$, and $3_{2,1}$--$2_{2,0}$ are used to derive the gas temperature through line ratio analysis. Dense cores are identified and traced using \tcsff{}, \dcntt{}, and \ntdptt{} emission.

\begin{figure}[ht]
        \centering
        \includegraphics[width=0.5\textwidth]{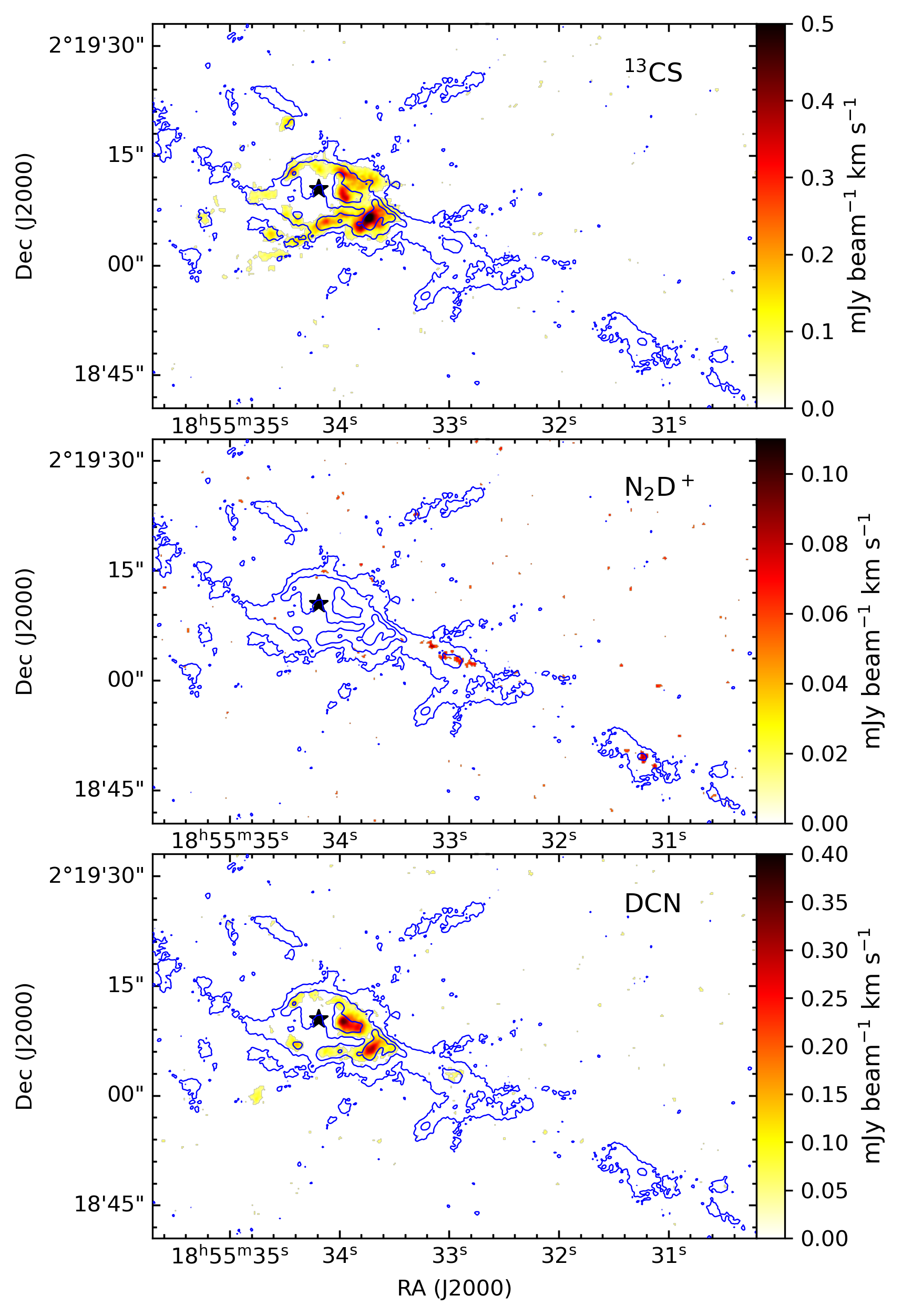}
        \caption{From top to bottom are the integrated intensity maps of \tcsff{}, \ntdptt{}, and \dcntt{}, integrated over 73--81~km~s$^{-1}$, 74--79~km~s$^{-1}$, and 72--81~km~s$^{-1}$, respectively. The blue contours show the 1.3~mm continuum emission with levels of $(3,15,45)\times$RMS, where the RMS is 0.11~Jy~beam$^{-1}$. The black star indicates the position of the \hii{} region.}
        \label{fig:mom0}
    \end{figure}

    From the integrated intensity maps of \tcsff{}, \ntdptt{}, and \dcntt{} in \autoref{fig:mom0}, we found that the spatial distribution of the emission of the three lines is different. The \ntdptt{} emission is confined to the filamentary structures located slightly farther away from the \hii{} region. In contrast, the \tcsff{} and \dcntt{} emission is detected almost exclusively around the \hii{} region. The reasons for these differences will be discussed in detail in \autoref{subsec:disc_mols}.

    \subsection{Core Identification}
    
\begin{figure*}[ht]
        \centering
        \includegraphics[width=1\textwidth,trim=0 1.5cm 2cm 2.5cm,clip]{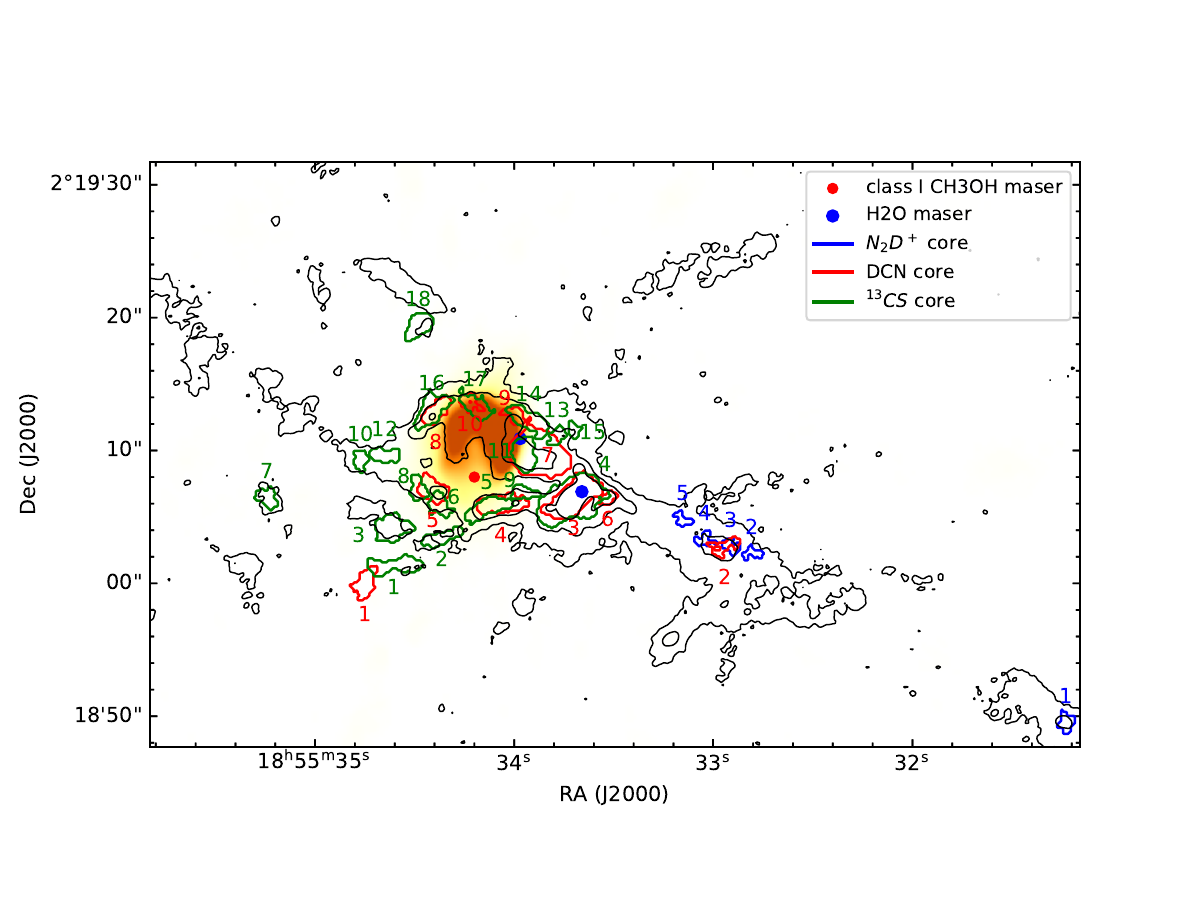}
        \caption{The background image shows the VLA 1.3~cm continuum. The black contours show the 1.3~mm continuum emission with levels of $(3, 15, 45)\times$RMS, where the RMS is 0.11~Jy~beam$^{-1}$. The red dots indicate maser detections extracted from the MaserDB database, while the blue dots show the H$_2$O masers detected in our VLA K-band observations.}
        \label{fig:identify}
    \end{figure*}

    We used the dendrogram algorithm implemented in the \textit{astrodendro} package \citep{Robitaille2019} to identify dense cores in the integrated intensity maps of the three molecular lines. The minimum flux density was set to 3$\sigma$, the minimum significance for the structures was 1$\sigma$, and the minimum area was set to the size of the synthesized beam. We considered leaves, the smallest structures in the hierarchy of the dendrogram, as dense cores. As shown in \autoref{fig:identify}, the spatial distribution of the identified structures is broadly consistent with the continuum cores reported by \citet{Cheng2024}, although the identified structures appear to cover fewer locations. Therefore, we adopt this identification of dendrogram leaves as dense cores in the following analysis. 
    
    We caution that certain tracers, such as \tcsff{}, may also be enhanced in shocked regions \citep[e.g.,][]{Liu2012_2,Minh2016}, potentially leading to misclassification. However, all of the \tcsff{} cores are spatially coincident with 1.3~mm continuum emission as shown in \autoref{fig:identify}, suggesting that they are likely associated with dense gas rather than shocks.
    
    We did not use the 1.3~mm continuum emission to identify dense cores, as the continuum emission near the \hii{} region may include contributions from free-free emission in addition to thermal dust emission, making it less suitable as a uniform tracer of dense gas across the entire region.

    \autoref{fig:identify} shows the results of the core identification. A total of 18, 10, and 5 cores were identified from the integrated intensity maps of \tcsff{}, \dcntt{}, and \ntdptt{}, respectively. The parameters of each core are listed in \autoref{tab:coreprop}. In this work, cores identified in different tracers are analyzed separately, and therefore the possible spatial overlap between tracers does not affect the subsequent analysis. To examine whether the dense cores are kinematically associated with the surrounding filamentary envelope, we projected the positions of all identified cores onto the filament skeleton and overlaid their centroid velocities on the NH$_3$ (1,1) PV diagrams shown in \autoref{fig:PV diagrams}. Except for the cores located close to the \hii{} region, the centroid velocities of the dense cores are generally consistent with the local NH$_3$ velocity structure. This agreement indicates that the dense cores are kinematically associated with the surrounding filamentary envelope and form part of the same coherent velocity structure.

    \begin{table*}
        \centering
        \caption{Core Properties Extracted by \textit{astrodendro}}\label{tab:coreprop}
        \resizebox{\textwidth}{!}{
        \begin{tabular}{cccccccccccc}
        \hline\hline
        ID & R.A.~(J2000) & Dec.~(J2000) & FWHM$^{a}$ & PA & Flux & $T$ & $\sigma_{\rm tot}$ & Mass & n & $M_{\rm vir}$ & $\alpha$ 
        \\ & (deg) & (deg) & ($10^3$ AU $\times$ $10^3$ AU) & (deg) & (mJy) & (K) & (km s$^{-1}$) & ($M_\odot$) & $\rm 10^5\ cm^{-3}$ & ($M_\odot$) & 
        \\ (1) & (2) & (3) & (4) & (5) & (6) & (7) & (8) & (9) & (10) & (11) & (12) \\
        \hline

        \multicolumn{12}{c}{N$_2$D$^+$} \\
        \hline
        1 & 283.8801 & 2.3138 & 4.3$\times$3.2 & 88.8 & 4.0$\pm$0.4 & 14.4$^{b}$ & 0.6$\pm$0.12 & $2.9^{+0.8}_{-0.6}$ & $35.9^{+13.1}_{-9.5}$ &$6.1^{+1.8}_{-1.6}$ & $2.1^{+0.9}_{-0.6}$ \\
        2 & 283.8867 & 2.3173 & 4.5$\times$2.5 & $-178.7$ & 1.1$\pm$0.1 & 17.1$\pm$0.9 & 0.50$\pm$0.37 & $0.6^{+0.2}_{-0.1}$ & $14.7^{+5.4}_{-3.9}$ & $3.5^{+4.3}_{-2.6}$ & $5.5^{+7.4}_{-4.2}$ \\
        3 & 283.8871 & 2.3174 & 4.1$\times$2.8 & 145.2 & 3.6$\pm$0.4 & 18.5$\pm$0.8 & 0.63$\pm$0.15 & $1.8^{+0.5}_{-0.4}$ & $32.5^{+11.9}_{-8.7}$ & $6.2^{+2.2}_{-1.9}$ & $3.3^{+1.6}_{-1.1}$ \\
        4 & 283.8877 & 2.3176 & 4.2$\times$3.1 & $-173.0$ & 3.2$\pm$0.3 & 16.3$\pm$0.7 & 0.50$\pm$0.13 & $1.9^{+0.6}_{-0.4}$ & $38.5^{+14.2}_{-10.3}$ & $3.7^{+1.4}_{-1.2}$ & $1.9^{+0.9}_{-0.6}$ \\
        5 & 283.8881 & 2.3180 & 4.2$\times$2.4 & 147.4 & 1.8$\pm$0.2 & 14.4$\pm$0.6 & 0.52$\pm$0.10 & $1.3^{+0.4}_{-0.3}$ & $31.7^{+11.7}_{-8.5}$ & $3.8^{+1.1}_{-0.9}$ & $2.9^{+1.2}_{-0.9}$ \\

        \hline
        \multicolumn{12}{c}{DCN} \\
        \hline
        1 & 283.8948 & 2.3167 & 6.9$\times$3.7 & 65.8 & 0.5$\pm$0.1 & 108.2$\pm$4.6 & \ldots & $0.03^{+0.01}_{-0.01}$ & $0.1^{+0.1}_{-0.0}$ & \ldots & \ldots \\
        2 & 283.8873 & 2.3174 & 7.2$\times$3.3 & $-159.9$ & 5.9$\pm$0.6 & 47.6$\pm$2.3 & 0.59$\pm$0.21 & $1.0^{+0.3}_{-0.2}$ & $9.1^{+3.3}_{-2.4}$ & $6.6^{+3.5}_{-2.8}$ & $6.7^{+4.2}_{-2.9}$ \\
        3 & 283.8904 & 2.3185 & 11.9$\times$5.6 & $-143.0$ & 109.7$\pm$11.0 & 201.7$\pm$21.2 & 1.23$\pm$0.08 & $3.8^{+1.2}_{-0.9}$ & $3.2^{+1.2}_{-0.9}$ & $64.2^{+6.8}_{-6.1}$ & $16.6^{+5.3}_{-4.0}$ \\
        4 & 283.8919 & 2.3183 & 10.7$\times$3.6 & $-174.9$ & 20.5$\pm$2.1 & 42.4$\pm$0.5 & 0.92$\pm$0.09 & $3.8^{+1.1}_{-0.8}$ & $8.7^{+3.2}_{-2.3}$ & $25.8^{+3.7}_{-3.4}$ & $6.7^{+2.2}_{-1.7}$ \\
        5 & 283.8934 & 2.3186 & 6.6$\times$4.2 & 147.0 & 5.2$\pm$0.5 & 42.6$\pm$0.9 & 0.70$\pm$0.06 & $1.0^{+0.3}_{-0.2}$ & $3.2^{+1.2}_{-0.8}$ & $13.2^{+1.6}_{-1.5}$ & $13.8^{+4.4}_{-3.3}$ \\
        6 & 283.8897 & 2.3184 & 4.0$\times$2.2 & $-149.4$ & 11.9$\pm$1.2 & 263.8$\pm$39.3 & 1.39$\pm$0.14 & $0.3^{+0.1}_{-0.1}$ & $7.4^{+2.8}_{-2.0}$ & $27.1^{+4.0}_{-3.7}$ & $84.1^{+29.1}_{-21.5}$ \\
        7 & 283.8912 & 2.3194 & 12.1$\times$8.2 & 160.3 & 95.8$\pm$9.6 & 70.6$\pm$1.1 & 0.83$\pm$0.05 & $10.1^{+3.0}_{-2.2}$ & $4.4^{+1.6}_{-1.2}$ & $36.5^{+3.8}_{-3.5}$ & $3.6^{+1.1}_{-0.9}$ \\
        8 & 283.8933 & 2.3203 & 7.2$\times$3.1 & 46.0 & 11.6$\pm$1.2 & 71.5$\pm$1.7 & 0.76$\pm$0.05 & $1.2^{+0.4}_{-0.3}$ & $5.5^{+2.0}_{-1.5}$ & $14.1^{+1.5}_{-1.3}$ & $11.6^{+3.6}_{-2.8}$ \\
        9 & 283.8917 & 2.3202 & 5.0$\times$3.0 & 147.0 & 7.0$\pm$0.7 & 73.9$\pm$2.1 & 1.08$\pm$0.10 & $0.7^{+0.2}_{-0.2}$ & $7.5^{+2.7}_{-2.0}$ & $21.2^{+3.0}_{-2.7}$ & $30.2^{+9.9}_{-7.5}$ \\
        10 & 283.8926 & 2.3204 & 4.9$\times$3.5 & 149.5 & \ldots & 69.9$\pm$2.5 & 0.81$\pm$0.10 & \ldots & \ldots & $11.9^{+2.1}_{-1.9}$ & \ldots \\

        \hline
        \multicolumn{12}{c}{\tcs{}} \\
        \hline
        1 & 283.8942 & 2.3170 & 12.6$\times$3.2 & $-176.0$ & 1.0$\pm$0.1 & 36.7$\pm$1.0 & 0.60$\pm$0.06 & $0.2^{+0.1}_{-0.0}$ & $0.6^{+0.2}_{-0.2}$ & $10.3^{+1.5}_{-1.4}$ & $47.1^{+15.7}_{-11.8}$ \\
        2 & 283.8932 & 2.3176 & 9.2$\times$2.6 & $-161.9$ & 1.1$\pm$0.1 & 35.6$\pm$0.9 & 0.62$\pm$0.02 & $0.2^{+0.1}_{-0.1}$ & $1.3^{+0.5}_{-0.3}$ & $8.8^{+0.7}_{-0.7}$ & $35.7^{+10.9}_{-8.4}$ \\
        3 & 283.8942 & 2.3178 & 6.7$\times$5.1 & $-170.4$ & 2.8$\pm$0.3 & 30.8$\pm$1.2 & 0.53$\pm$0.05 & $0.7^{+0.2}_{-0.2}$ & $1.8^{+0.7}_{-0.5}$ & $8.6^{+1.2}_{-1.1}$ & $11.4^{+3.8}_{-2.8}$ \\
        4 & 283.8905 & 2.3184 & 14.0$\times$7.8 & $-152.1$ & 141.6$\pm$14.2 & 157.2$\pm$11.5 & 1.11$\pm$0.04 & $6.4^{+1.9}_{-1.4}$ & $2.6^{+1.0}_{-0.7}$ & $66.4^{+5.0}_{-4.9}$ & $10.3^{+3.2}_{-2.4}$ \\
        5 & 283.8922 & 2.3182 & 10.2$\times$3.5 & $-158.7$ & 13.2$\pm$1.3 & 43.1$\pm$0.5 & 0.76$\pm$0.02 & $2.4^{+0.7}_{-0.5}$ & $7.1^{+2.6}_{-1.9}$ & $16.1^{+1.2}_{-1.1}$ & $6.7^{+1.9}_{-1.6}$ \\
        6 & 283.8932 & 2.3183 & 5.4$\times$3.8 & 125.7 & 1.7$\pm$0.2 & 38.1$\pm$0.8 & 0.74$\pm$0.05 & $0.3^{+0.1}_{-0.1}$ & $2.3^{+0.8}_{-0.6}$ & $11.7^{+1.2}_{-1.1}$ & $33.7^{+10.3}_{-8.1}$ \\
        7 & 283.8968 & 2.3185 & 4.9$\times$3.5 & 140.1 & 1.6$\pm$0.2 & 46.5$\pm$1.7 & 0.58$\pm$0.03 & $0.3^{+0.1}_{-0.1}$ & $2.2^{+0.8}_{-0.6}$ & $6.7^{+0.7}_{-0.6}$ & $25.8^{+7.9}_{-6.1}$ \\
        8 & 283.8937 & 2.3186 & 5.3$\times$3.2 & 103.0 & 2.1$\pm$0.2 & 43.9$\pm$1.0 & 0.55$\pm$0.11 & $0.4^{+0.1}_{-0.1}$ & $3.2^{+1.2}_{-0.8}$ & $6.0^{+1.8}_{-1.6}$ & $15.4^{+6.7}_{-4.8}$ \\
        9 & 283.8915 & 2.3186 & 5.9$\times$2.7 & 174.8 & 5.4$\pm$0.5 & 45.4$\pm$0.8 & 0.65$\pm$0.03 & $0.9^{+0.3}_{-0.2}$ & $9.3^{+3.4}_{-2.5}$ & $7.7^{+0.7}_{-0.6}$ & $8.4^{+2.5}_{-2.0}$ \\
        10 & 283.8949 & 2.3192 & 4.4$\times$3.0 & 92.4 & 1.8$\pm$0.2 & 51.4$\pm$1.9 & 0.72$\pm$0.07 & $0.3^{+0.1}_{-0.1}$ & $3.7^{+1.3}_{-1.0}$ & $8.7^{+1.3}_{-1.2}$ & $31.8^{+10.8}_{-7.9}$ \\
        11 & 283.8915 & 2.3194 & 7.1$\times$4.5 & 97.6 & 37.9$\pm$3.8 & 76.4$\pm$1.5 & 0.92$\pm$0.03 & $3.7^{+1.1}_{-0.8}$ & $10.4^{+3.8}_{-2.7}$ & $24.2^{+1.8}_{-1.7}$ & $6.6^{+2.0}_{-1.5}$ \\
        12 & 283.8944 & 2.3193 & 6.6$\times$2.9 & 175.5 & 2.4$\pm$0.2 & 45.7$\pm$1.6 & 0.66$\pm$0.08 & $0.4^{+0.1}_{-0.1}$ & $3.0^{+1.1}_{-0.8}$ & $9.1^{+1.6}_{-1.4}$ & $21.9^{+7.6}_{-5.8}$ \\
        13 & 283.8908 & 2.3197 & 4.6$\times$2.8 & 48.1 & 4.7$\pm$0.5 & 72.6$\pm$1.9 & 0.77$\pm$0.02 & $0.5^{+0.1}_{-0.1}$ & $6.7^{+2.5}_{-1.8}$ & $10.0^{+0.7}_{-0.7}$ & $20.6^{+6.1}_{-4.8}$ \\
        14 & 283.8914 & 2.3201 & 9.7$\times$3.8 & 147.9 & 14.4$\pm$1.4 & 65.3$\pm$1.2 & 0.76$\pm$0.02 & $1.6^{+0.5}_{-0.4}$ & $4.2^{+1.5}_{-1.1}$ & $17.2^{+1.2}_{-1.2}$ & $10.4^{+3.1}_{-2.4}$ \\
        15 & 283.8904 & 2.3199 & 3.5$\times$2.4 & 107.9 & 2.2$\pm$0.2 & 57.0$\pm$1.9 & 0.90$\pm$0.04 & $0.3^{+0.1}_{-0.1}$ & $7.7^{+2.8}_{-2.0}$ & $11.0^{+0.9}_{-0.9}$ & $37.1^{+11.2}_{-8.8}$ \\
        16 & 283.8934 & 2.3203 & 8.5$\times$4.3 & 46.9 & 14.6$\pm$1.5 & 65.2$\pm$1.4 & 0.74$\pm$0.03 & $1.7^{+0.5}_{-0.4}$ & $3.8^{+1.4}_{-1.0}$ & $16.6^{+1.3}_{-1.2}$ & $9.9^{+3.0}_{-2.4}$ \\
        17 & 283.8925 & 2.3204 & 7.5$\times$3.5 & 141.8 & $\ldots$ & 53.3$\pm$1.9 & 0.82$\pm$0.06 & $\ldots$ & \ldots & $16.9^{+2.0}_{-1.8}$ & $\ldots$ \\
        18 & 283.8936 & 2.3221 & 6.3$\times$3.7 & $-136.9$ & 1.1$\pm$0.1 & 78.7$\pm$4.6 & 0.76$\pm$0.03 & $0.1^{+0.0}_{-0.0}$ & $0.4^{+0.2}_{-0.1}$ & $14.4^{+1.2}_{-1.1}$ & $133.3^{+40.8}_{-31.3}$ \\
        \hline\hline
        \end{tabular}
        }
        \tablefoot{ 
            $a$: Full Width at Half Maximum of the dense core. For DCN core~1, the signal-to-noise ratio is insufficient for reliable spectral fitting. 
            For DCN core~10 and \tcs{} core~17, the 1.3~mm continuum emission becomes negligible after subtracting the free--free emission, and no reliable mass can be derived. 
            $b$: For N$_2$D$^+$ Core 1, the temperature was fixed to the minimum temperature measured among the other five N$_2$D$^+$ cores. Consequently, no temperature uncertainty is available for this core.}
    \end{table*}

    \subsection{Physical Properties of the Dense Cores}\label{subsec:results_cores}

    \begin{figure}[ht]
        \centering
        \includegraphics[width=0.5\textwidth]{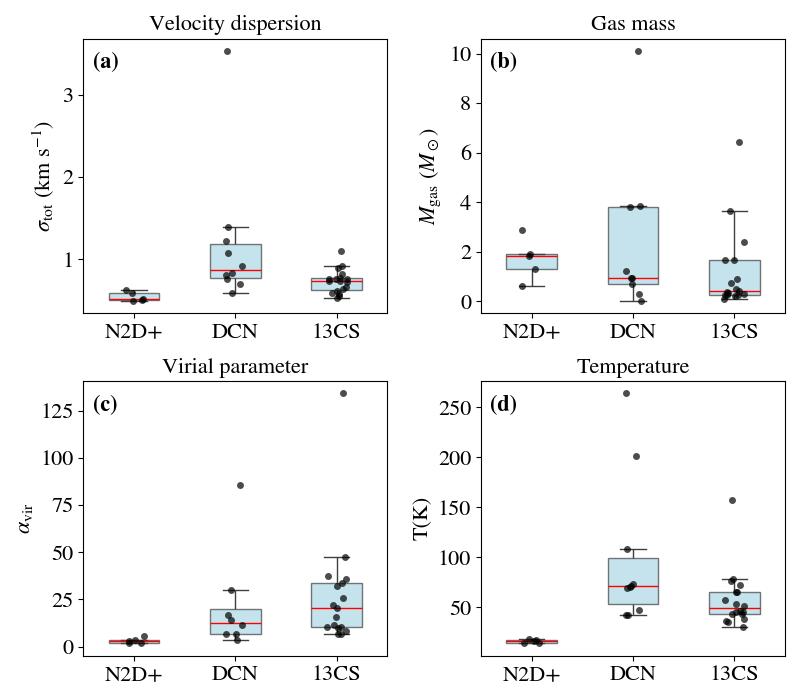}
        \caption{Panels a to d show the distributions of the total velocity dispersion, gas mass, virial parameter, and temperature for gas traced by \ntdp{}, DCN, and \tcs{}. In each panel, black points represent individual cores, while the boxplots summarize the corresponding distributions. The red line indicates the median value of each distribution.}
        \label{fig:Properties}
    \end{figure}

        \subsubsection{Temperatures of the Dense Cores}\label{subsubsec:results_temp}
        Assuming local thermodynamic equilibrium (LTE) conditions, we estimated the temperature of the cores traced by DCN and \tcs{} using the FFTL code\footnote{\url{https://github.com/xinglunju/FFTL}} \citep[e.g.,][]{Zhang2025} by fitting the three $\mathrm{H_2CO}$ transitions $3_{0,3}$--$2_{0,2}$, $3_{2,2}$--$2_{2,1}$, and $3_{2,1}$--$2_{2,0}$, which are sensitive to the gas temperature \citep[e.g.,][]{Mangum1993}. For the cores traced by \ntdptt{}, where the $\mathrm{H_2CO}$ lines have relatively low signal-to-noise ratios, the temperature was estimated by fitting the $\mathrm{NH_3}$ (1,1) and (2,2) lines using the pyAmor code\footnote{\url{https://github.com/xinglunju/pyAmor}} \citep[e.g.,][]{Lu2018}. The results are presented in \autoref{tab:coreprop}. 
        
        To assess the consistency between the temperatures derived from different tracers, we list the results in \autoref{tab:temperature}. For most cores, the temperature differences fall within the expected systematic uncertainties \citep{Li2026}. However, deviations are found for DCN cores~3 and~6, as well as \tcs{} core~4, where the discrepancies exceed the typical systematic level. These outliers are spatially clustered near the edge of the \hii{} region shell and also show relatively enhanced velocity dispersions compared to the rest of the sample, suggesting stronger dynamical activity. We will discuss these cores in more detail in \autoref{subsec:OutlierCores}

        As an additional consistency check, \rev{we compared the physical properties of dense cores for which both NH$_3$- and H$_2$CO-derived temperatures are available, including two \ntdp{} cores, three DCN cores, and five \tcs{} cores. For the \ntdp{} cores, adopting H$_2$CO-based temperatures decreases the gas masses by an average of 69\%, while increasing the velocity dispersions by 13\% and the virial parameters by a factor of 3.1 on average. In contrast, for the DCN cores, adopting NH$_3$-based temperatures increases the gas masses by a factor of 9.7 on average, while decreasing the velocity dispersions and virial parameters by 19\% and 83\%, respectively. For the \tcs{} cores, the gas masses increase by a factor of 2.9 on average, while the velocity dispersions and virial parameters decrease by 8.5\% and 41\%, respectively. However, this approach is unlikely to be physically appropriate, as different molecular tracers may probe different gas components along the line of sight. In particular, \ntdp{} preferentially traces cold, dense gas (see \autoref{subsec:disc_mols} for a more detailed discussion), whereas $\mathrm{H_2CO}$ may probe warmer material along the same line of sight rather than the cold core itself. Therefore, we retained the tracer-specific temperature assumptions adopted in our analysis.}

        \begin{table}
        \centering
        \caption{Temperature comparison derived from NH$_3$ and H$_2$CO.}\label{tab:temperature}
        \begin{tabular}{lccc}
        \hline
        Core & $\mathrm{NH_3}$ (K) & $\mathrm{H_2CO}$ (K) & $\Delta T$ (K) \\
        \hline

        \ntdp{} core 1 & undetected & undetected & $\ldots$ \\
        \ntdp{} core 2 & 17.1$\pm$0.9 & undetected & $\ldots$ \\
        \ntdp{} core 3 & 18.5$\pm$0.8 & undetected & $\ldots$ \\
        \ntdp{} core 4 & 16.3$\pm$0.7 & 39.9$\pm$2.0 & 23.6 \\
        \ntdp{} core 5 & 14.4$\pm$0.6 & 35.3$\pm$2.7 & 21.1 \\

        DCN core 1 & undetected & 108.2$\pm$4.6 & $\ldots$ \\
        DCN core 2 & 18.6$\pm$0.7 & 47.6$\pm$2.3 & 29.0 \\
        DCN core 3 & 23.5$\pm$1.2 & 201.7$\pm$21.2 & 178.2 \\
        DCN core 4 & unreliable fit & 42.4$\pm$0.5 & $\ldots$ \\
        DCN core 5 & undetected & 42.6$\pm$0.9 & $\ldots$ \\
        DCN core 6 & 25.4$\pm$2.5 & 263.8$\pm$39.3 & 238.4 \\
        DCN core 7 & undetected & 70.6$\pm$1.1 & $\ldots$ \\
        DCN core 8 & undetected & 71.5$\pm$1.7 & $\ldots$ \\
        DCN core 9 & undetected & 73.9$\pm$2.1 & $\ldots$ \\
        DCN core 10 & undetected & 69.9$\pm$2.5 & $\ldots$ \\

        \tcs{} core 1 & undetected & 36.7$\pm$1.0 & $\ldots$ \\
        \tcs{} core 2 & unreliable fit & 35.6$\pm$0.9 & $\ldots$ \\
        \tcs{} core 3 & 24.7$\pm$3.7 & 30.8$\pm$1.2 & 6.1 \\
        \tcs{} core 4 & 23.5$\pm$1.1 & 157.2$\pm$11.5 & 133.7 \\
        \tcs{} core 5 & unreliable fit & 43.1$\pm$0.5 & $\ldots$ \\
        \tcs{} core 6 & unreliable fit & 38.1$\pm$0.8 & $\ldots$ \\
        \tcs{} core 7 & undetected & 46.5$\pm$1.7 & $\ldots$ \\
        \tcs{} core 8 & 24.3$\pm$3.7 & 43.9$\pm$1.0 & 19.6 \\
        \tcs{} core 9 & 26.8$\pm$3.7 & 45.4$\pm$0.8 & 18.6 \\
        \tcs{} core 10 & 51.9$\pm$8.6 & 51.4$\pm$1.9 & 0.5 \\
        \tcs{} core 11 & undetected & 76.4$\pm$1.5 & $\ldots$ \\
        \tcs{} core 12 & undetected & 45.7$\pm$1.6 & $\ldots$ \\
        \tcs{} core 13 & unreliable fit & 72.6$\pm$1.9 & $\ldots$ \\
        \tcs{} core 14 & undetected & 65.3$\pm$1.2 & $\ldots$ \\
        \tcs{} core 15 & undetected & 57.0$\pm$1.9 & $\ldots$ \\
        \tcs{} core 16 & undetected & 65.2$\pm$1.4 & $\ldots$ \\
        \tcs{} core 17 & undetected & 53.3$\pm$1.9 & $\ldots$ \\
        \tcs{} core 18 & undetected & 78.8$\pm$4.6 & $\ldots$ \\

        \hline
        \end{tabular}
        \tablefoot{
        Values labeled as ``undetected'' indicate that the peak intensity is below 3$\sigma$. ``Unreliable fit'' denotes spectra where some transitions are not detected above 3$\sigma$, resulting in insufficient constraints for a reliable fit. For N$_2$D$^+$ Core 1, no reliable temperature could be derived from either the $\mathrm{H_2CO}$ or $\mathrm{NH_3}$ data owing to their low signal-to-noise ratios. We therefore adopted the minimum temperature measured among the other five N$_2$D$^+$ cores as a conservative estimate.
        }
        \end{table}

\begin{figure}[ht]
            \centering
            \includegraphics[width=0.5\textwidth]{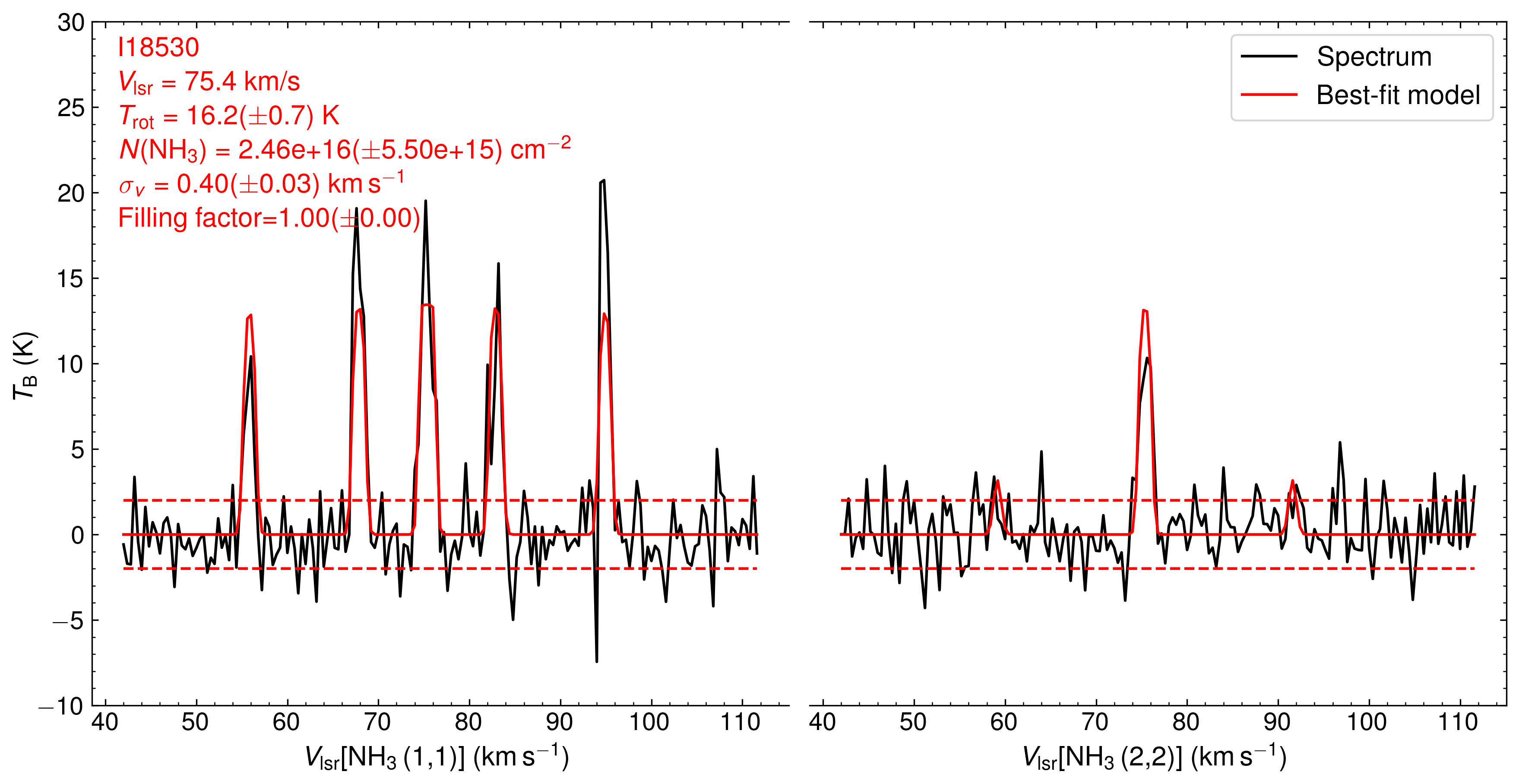}
            \caption{Fitting results of $\mathrm{NH_3\ (1,1)}$ and (2,2) spectra toward \ntdp{} core 4. The red-shifted outer satellite line of the $\mathrm{NH_3\ (1,1)}$ transition is significantly stronger than the main line.}
            \label{fig:temperature}
        \end{figure}

        When estimating the temperature for the \ntdp{} cores using $\mathrm{NH_3}$ lines, we found that in several cases the assumption of LTE may not hold. In particular, \ntdp{} cores 4 and 5 exhibit anomalous $\mathrm{NH_3\ (1,1)}$ line profiles, where the red-shifted outer satellite lines are stronger than the main component (\autoref{fig:temperature}). Such a feature indicates that the excitation conditions in these regions deviate from LTE. In the non-LTE models of \citet{Stutzki1985}, significant $\mathrm{NH_3}$ hyperfine anomalies are predicted under physical conditions similar to those of \ntdp{} cores 4 and 5, which have sizes of $\sim10^{-2}$~pc and $\mathrm{H_2}$ volume densities of $\sim10^{6}~\mathrm{cm^{-3}}$. This suggests that an exceptionally high $\mathrm{NH_3}$ column density may be responsible for the observed hyperfine anomalies.

        \subsubsection{Masses of the Dense Cores}
        We estimated the gas mass of the cores with the ALMA 1.3~mm continuum. For each core, the 1.3 mm flux was measured by integrating the continuum emission within the corresponding dendrogram boundaries.

\begin{figure}[ht]
            \centering
            \includegraphics[width=0.5\textwidth]{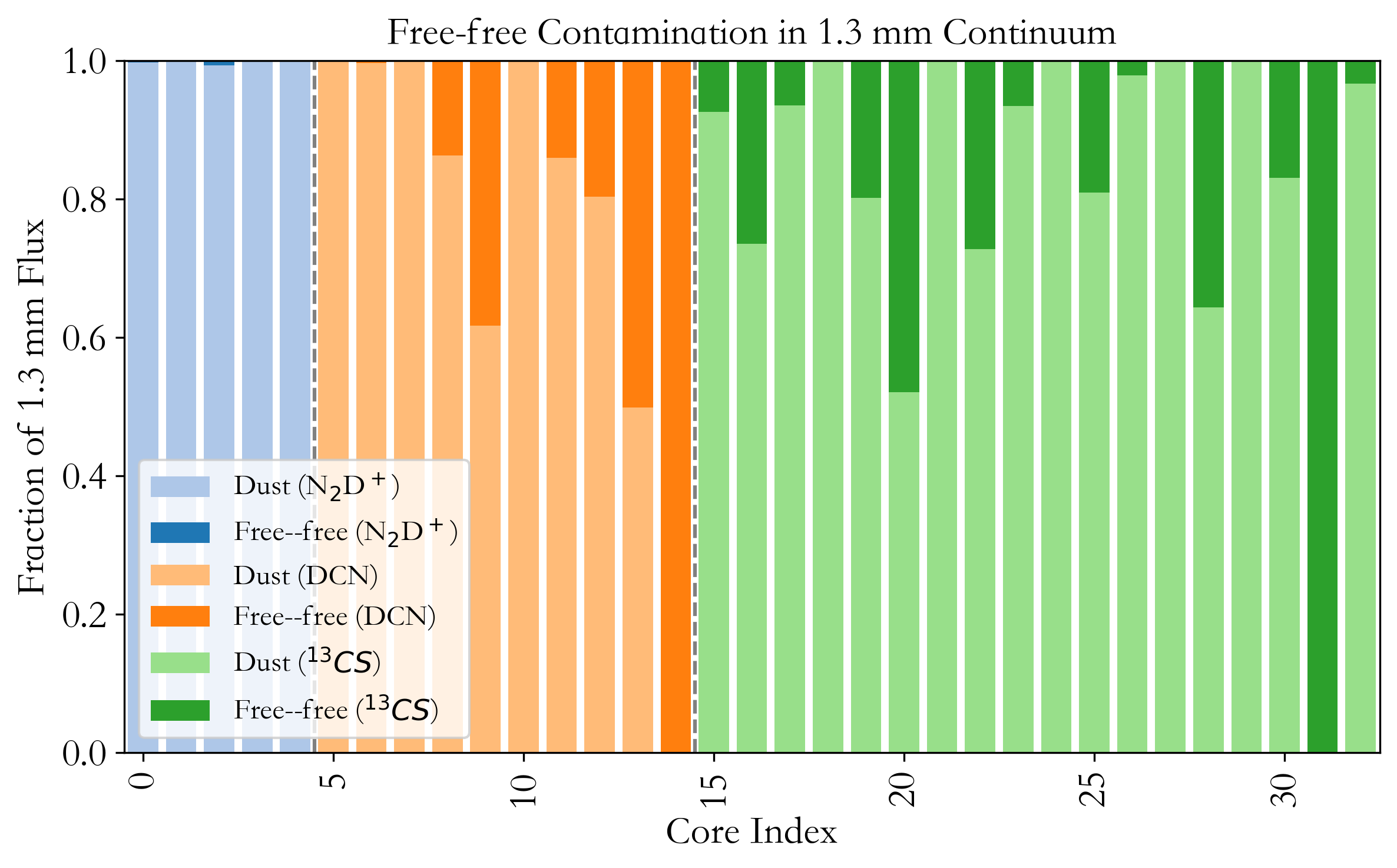}
            \caption{Fractional contributions of dust emission and free--free emission to the total 1.3~mm continuum flux for all identified cores. The free--free component is estimated from the 13~mm flux, while the dust emission is obtained by subtracting the 13~mm flux from the 1.3~mm flux. Each bar represents an individual core, with the lower (lighter) portion indicating the dust contribution and the upper (darker) portion indicating the free--free contribution. The cores are grouped into three categories (N$_2$D$^+$, DCN, and $^{13}$CS), separated by vertical dashed lines.}
            \label{fig:FluxFraction}
        \end{figure}

        The 1.3~mm continuum is often dominated by thermal dust emission. As dust emission decreases rapidly with spectral index of $\sim$3.8 towards centimeter wavelength \citep{Kohler2015}, 1.3~cm continuum is primarily attributed to free-free emission. Therefore, we adopted the following two assumptions: 1) the 1.3~cm continuum originates entirely from free-free emission with a spectral index of $-$0.1 \citep{Condon1992}; and 2) the thermal dust emission is assumed to be optically thin at 1.3~mm, which is supported by the fact that the peak brightness temperature is much lower than the dust temperature. Based on the spectral index of $-$0.1, we derived the free-free contribution at 1.3~mm and subtracted it to isolate the pure thermal dust emission. The angular resolution of the VLA 1.3 cm data is slightly coarser than that of the ALMA 1.3 mm continuum. However, since the free–free contribution was estimated using the integrated flux density of each core region, the difference in angular resolution does not significantly affect the subtraction. The resulting dust and free-free fractions for each core are shown in \autoref{fig:FluxFraction}.

        Subsequently, the dust mass can be determined via the following equation:
        \begin{equation}
            M_{\text{dust}} = \frac{d^2 F_\nu}{\kappa_\nu B_\nu(T_{\text{dust}})},
            \label{eq:dust_mass}  
        \end{equation}
        where $d$ is the distance to the source, $F_\nu$ is the flux of the thermal dust emission, $\kappa_\nu$ is the dust opacity at the observed frequency $\nu$. We adopted $\kappa_\text{1.3~mm}$ = 0.9~$\mathrm{cm^2\,g^{-1}}$, corresponding to conditions at a density of $\mathrm{10^6~cm^{-3}}$ for thin ice mantles \citep{Ossenkopf1994}. $T_\text{dust}$ is the dust temperature, which is assumed to be equal to the gas temperature derived in \autoref{subsubsec:results_temp} given the high densities in the dense cores \citep{Goldsmith2001}. $B_\nu$ is the Planck function at the dust temperature $T_\text{dust}$. Assuming a gas-to-dust mass ratio $R_{\text{gd}}$ of 100, we then derived the gas mass of each core from the dust mass:
        \begin{equation}
            M_{\text{gas}} = R_{\text{gd}}\times M_{\text{dust}}.
            \label{eq:dust_mass}  
        \end{equation}

        The derived gas masses for all cores are listed in \autoref{tab:coreprop}. The uncertainties of these quantities were estimated by propagating the errors of the input parameters using a Monte Carlo approach. Specifically, we assumed a relative uncertainty of $10\%$ for the distance (and thus the radius), $10\%$ for the flux density $F_\nu$ (see the ALMA Technical Handbook), $23\%$ for the dust opacity $\kappa_\nu$, and $28\%$ for the gas-to-dust mass ratio $R_{\rm gd}$ \citep[e.g.,][]{Sanhueza2017,Xu2024,Shen2024}. All parameters were assumed to follow uniform distributions within their respective uncertainty ranges. We then generated $10^4$ random realizations to derive the resulting distributions of the physical quantities. The final uncertainties are defined by the 16th and 84th percentiles of these distributions, corresponding to the lower and upper bounds, respectively. The uncertainties reported in columns~9--12 were derived following this procedure.

        Additionally, to evaluate the amount of missing flux, we compared the integrated flux densities measured from the original and feathered continuum images within the dendrogram boundaries of each core. We find that the typical missing flux fraction is approximately 57\%, indicating that a substantial fraction of the large scale diffuse emission is filtered out by the interferometric observations. However, since the dense cores are compact structures, the interferometric data are still expected to recover the compact emission reliably. In contrast, the additional flux recovered in the feathered image is likely dominated by diffuse emission surrounding the cores. Therefore, we adopted the original ALMA continuum image for the core mass estimates, while the feathered image was used only for more extended structures such as the \hii{} shell.
        
        The median core masses show a tentative increasing trend from \tcs{} to DCN to \ntdp{} (\autoref{fig:Properties}). This result contrasts with previous studies \citep[e.g.,][]{Cunningham2023}, which found that DCN and \tcs{} tend to trace higher-mass cores, while \ntdp{} is mainly associated with lower or intermediate-mass cores. This discrepancy is likely related to the temperature estimates adopted in this work. As demonstrated in the consistency check presented in \autoref{subsubsec:results_temp}, using $\mathrm{H_2CO}$-based temperatures for the same \ntdp{} cores leads to systematically higher temperatures and consequently lower derived masses. In contrast, \tcs{} and DCN core temperatures are derived from $\mathrm{H_2CO}$, whereas \ntdp{} cores are characterized using $\mathrm{NH_3}$-based temperatures. This inconsistency in temperature tracers therefore likely introduces a systematic bias in the derived core masses.

        \subsubsection{Velocity Dispersions of the Dense Cores}
        
        For the three populations of cores traced by \tcs{}, DCN, and \ntdp{}, we fitted the corresponding core-averaged molecular line spectra to obtain the observed velocity dispersions $\mathrm{\sigma_{obs}}$, and calculated the total velocity dispersion using the following method: 
        \begin{equation}
            \sigma_{\rm int}=\sqrt{\sigma_{\rm obs}^2-(\Delta_{\rm ch}/(2\sqrt{2\ln2}))^2},
            \label{eq:velocity dispersions observed}  
        \end{equation}
        where $\sigma_{\rm int}$ is the intrinsic line-of-sight velocity dispersion, defined as the Gaussian sigma of the fitted line profile after correcting for the channel width. In our data, the channel width $\Delta_{\rm ch}$ is 0.35~\kms{}. 
        
        To derive the velocity dispersion of the bulk molecular gas in the cores, characterized by the mean molecular weight of the gas, rather than that of the observed molecular species, we corrected the observed line widths for thermal broadening. The non-thermal component in the velocity dispersion is
        \begin{equation}
            \sigma_{\rm nt}^2=\sigma_{\rm int}^2-\sigma_{\rm th}^2,
            \label{eq:non-thermal}  
        \end{equation}
        where $\sigma_{\rm th}^2=(k_{\rm B}T)/(\mu_m m_p)$, in which $k_{\rm B}$ is the Boltzmann constant, $T$ is the gas temperature, $\mu_m$ is the molecular weight of the observed species (e.g., $\mu_m$[\tcs{}]=45, $\mu_m$[DCN]=28, and $\mu_m$[\ntdp{}]=30), and $m_p$ is the proton mass. The thermal velocity dispersion of the molecular gas is
        \begin{equation}
            c_{\rm s}=\sigma_{\rm th,gas}=\sqrt{k_{\rm B}T/(\mu_m m_p)},
            \label{eq:thermal}  
        \end{equation}
        with $\mu_m=2.33$ for the molecular hydrogen. The total velocity dispersion of the cores is derived following
        \begin{equation}
            \sigma_{\rm tot}^2=\sigma_{\rm nt}^2+c_{\rm s}^2,
            \label{eq:tot}  
        \end{equation}
        and is presented in \autoref{tab:coreprop} and \autoref{fig:Properties}.

        \subsubsection{Equilibrium of the Dense Cores}
        
        We calculated the virial parameter for each core as a diagnostic of its dynamical equilibrium. The virial parameter can be expressed as the ratio of the virial mass to the gas mass,
        \begin{equation}
            M_{\rm vir}=\frac{5\sigma_{\rm tot}^2R_{\rm eff}}{G},  
        \end{equation}
        and
        \begin{equation}
            \alpha_{\rm vir}=M_{\rm vir}/M_{\rm gas},
            \label{eq:virial parameter}
        \end{equation}
        The effective radius is defined as $R_{\rm eff,c} = \sqrt{A/\pi}$, where $A$ is the projected area of the core. If $\alpha_{\rm vir}<2$, the core is gravitationally bound and tends to collapse, whereas if $\alpha_{\rm vir}>2$, the core is not bound and may expand. The results are also presented in \autoref{tab:coreprop}. We note that the virial parameters may be overestimated, as the embedded stellar mass within the cores is not taken into account. In addition, it is possible that some of the identified structures are transient or partly affected by projection effects, rather than representing true gravitationally bound structures. Even with these uncertainties in mind, we found that the \ntdp{} cores consistently exhibit lower virial parameters compared to the DCN and \tcs{} cores. In contrast, the virial parameters of the DCN and \tcs{} cores show a much wider range, with some cores having significantly higher values (\autoref{fig:Properties}). This diversity may be related to their proximity to the \hii{} region, where cores closer are likely to be affected by stellar feedback, influencing their dynamical equilibrium. We will discuss this effect in more detail in \autoref{subsec:hiiinfluence}.

\section{Discussion}\label{sec:disc}

    \subsection{Spatial Distributions of \ntdp{}, $\mathrm{DCN}$, and \tcs{}}\label{subsec:disc_mols}
    
    As shown in \autoref{fig:mom0}, the three molecular tracers exhibit different spatial distributions, indicating that they probe gas under different physical or chemical conditions. \rev{The excitation properties of these tracers, including their critical densities and effective excitation densities, are summarized in \autoref{tab:tracers}. Because the standard definitions of critical density do not account for radiative trapping, the effective excitation densities are typically 1–2 orders of magnitude lower than the critical densities \citep{Shirley2015}.}

    \ntdp{} is mainly formed through reactions between $\mathrm{N_2}$ and $\mathrm{H_2D^+}$, and is destroyed primarily by CO or electrons \citep{Turner2001,shanghuo2022}. CO tends to freeze out onto dust grains in cold regions; therefore the abundance of \ntdp{} increases under such conditions, whereas in warmer environments, where CO returns to the gas phase, \ntdp{} is efficiently destroyed and its emission becomes weaker \citep{Sakai2022}. 

    In contrast, DCN can be produced through both low-temperature and high-temperature formation channels, with the high-temperature pathway being the dominant mechanism overall \citep{Turner2001,Albertsson2013,Salinas2017}. At elevated temperatures, deuterium fractionation becomes less efficient due to the enhancement of backward reactions. However, DCN is a relatively stable molecule and is not rapidly destroyed under warmer conditions. In addition, thermal desorption can release DCN from dust grains into the gas phase, further enhancing its abundance \citep{Turner2001}. As a result, DCN is more readily detected in comparatively warmer regions \citep{Gerner2015,Salinas2017,Butterworth2024}. \tcs{} is also a commonly used tracer of dense gas in massive star-forming regions, although it may be enhanced by shocks as well \citep{Liu2015,Law2025}. Due to its relatively high excitation temperature (33~K), it traces comparatively warmer gas than \ntdp{}.

    \begin{table}
        \centering
        \caption{Properties of the dense core tracers \rev{at a kinetic temperature of 20~K}}
        \begin{tabular}{lccc}
        \hline
        Dense core tracer
        & \ntdptt{} 
        & \dcntt{} 
        & \tcsff{} \\
        \hline

        $E_{\rm u}/k$ (K)
        & 22 
        & 21  
        & 33  \\

        $n_{\rm cr}$ ($\mathrm{cm^{-3}}$) 
        & $1.9\times10^{6}$ 
        & $2.3\times10^{7}$ 
        & $3.7\times10^{6}$ \\

        $n_{\rm eff}$ ($\mathrm{cm^{-3}}$) 
        & $6.8\times10^{4}$ 
        & $7.3\times10^{4}$ 
        & $2.5\times10^{5}$ \\

        number of cores 
        & 5 
        & 10 
        & 18 \\

        \hline
        \end{tabular}
        \label{tab:tracers}
        \tablefoot{
        {The line data are taken from the Cologne Database for Molecular Spectroscopy \citep{Muller2001} and the Leiden Atomic and Molecular Database \citep{Schoier2005}. \rev{The effective excitation densities ($n_{\rm eff}$) are adopted from \citet{Shirley2015} and approximated using the corresponding main isotopologues.}}
}
    \end{table}

    To investigate whether the non-detection of DCN and \tcs{} in the dense cores farther away from the \hii{} region is caused by insufficient sensitivity, we estimated the minimum detectable column densities under the assumptions of local thermodynamic equilibrium (LTE) and optically thin emission.
    For DCN, the $4\sigma$ detection limit corresponds to a brightness temperature of approximately 0.68~K. For an order-of-magnitude estimate, we adopted the Rayleigh--Jeans approximation,
    \begin{equation}
            T_{\rm B} \approx T_{\rm ex}(1-e^{-\tau}),  
    \end{equation}
    which reduces to
    \begin{equation}
            T_{\rm B} \approx T_{\rm ex}\tau,
    \end{equation}
    under the optically thin condition ($\tau \ll 1$), where $T_{\rm B}$ is the brightness temperature, $T_{\rm ex}$ is the excitation temperature, and $\tau$ is the optical depth. Assuming an excitation temperature of $T_{\rm ex}=17$~K, representative of the quiescent dense cores far from the \hii{} region, we obtain an optical depth of $\tau \sim 0.04$.
    The total column density can then be estimated using
    \begin{equation}
            N_{\rm tot}=\frac{8\pi \nu^3}{c^3 A_{ul}}\frac{Q(T_{\rm ex})}{g_u}\frac{\exp(E_u/kT_{\rm ex})}{\exp(h\nu/kT_{\rm ex})-1}\tau \Delta v
    \end{equation}
    where $\nu$ is the transition frequency, $A_{ul}$ is the Einstein coefficient, $Q(T_{\rm ex})$ is the partition function, $g_u$ is the degeneracy of the upper level, $E_u$ is the upper-state energy. The molecular spectroscopic parameters were obtained from the Cologne Database for Molecular Spectroscopy (CDMS; \citealt{Mulle2001,Mulle2005}). Adopting a typical FWHM linewidth of $\Delta v = 1.25~\mathrm{km\,s^{-1}}$, measured from the \ntdp{} emission, we derive a minimum detectable DCN column density of $\sim2.0\times10^{12}~\mathrm{cm^{-2}}$.
    Applying the same procedure to \tcs{}, we derive a minimum detectable column density of $\sim5.7\times10^{12}~\mathrm{cm^{-2}}$.
    Assuming abundances of $X$(\tcs{}) $\sim5.4\times10^{-11}$ and $X({\rm DCN}) \sim8.3\times10^{-12}$ \citep{Li2022}, the corresponding minimum detectable H$_2$ column densities are estimated to be $\sim1.1\times10^{23}~\mathrm{cm^{-2}}$ and $\sim2.4\times10^{23}~\mathrm{cm^{-2}}$, respectively.

    The H$_2$ column densities derived from the dust continuum emission toward the \ntdp{} cores range from $\sim7\times10^{22}$ to $\sim2\times10^{23}~\mathrm{cm^{-2}}$. This range is comparable to the minimum detectable H$_2$ column densities inferred from the DCN and \tcs{} sensitivity limits. Considering the uncertainties in the adopted molecular abundances, the absence of detectable DCN and \tcs{} emission in the dense cores farther from the \hii{} region may be explained, at least in part, by the limited observational sensitivity.

    \subsection{Dynamical State of the \hii{} Region}\label{subsec:results_hii}

    \begin{figure}[ht]
        \centering
        \includegraphics[width=0.5\textwidth]{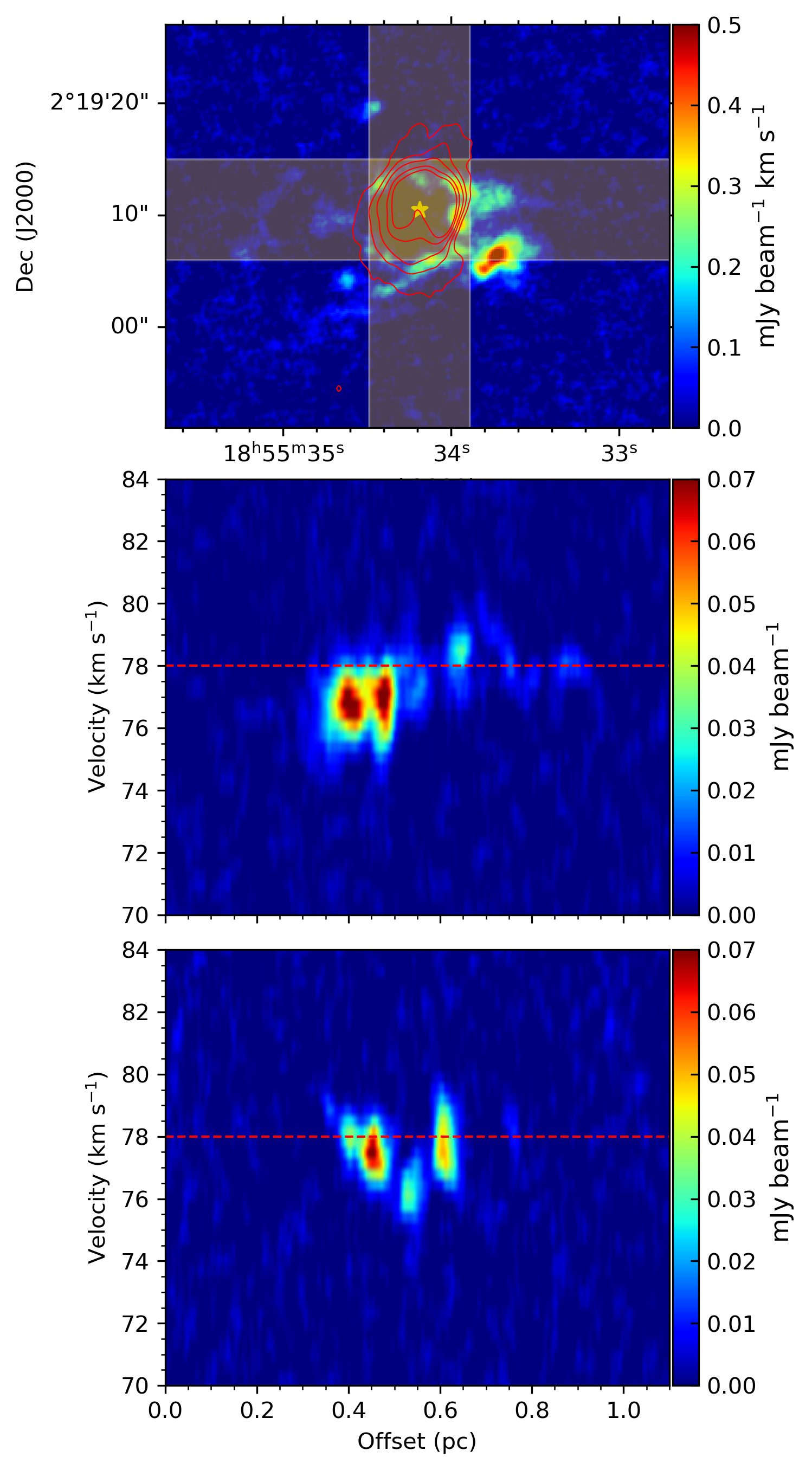}
        \caption{Top panel: Integrated intensity map of \tcs{}. The gold semi-transparent rectangles indicate the two paths along which the PV diagrams are extracted. The red contours show the 1.3~cm continuum emission at levels of $(12,50,100,200,300)\times$RMS, where the RMS is $1.5\times10^{-2}$~mJy~beam$^{-1}$. The yellow star marks the center of the \hii{} region. Middle and bottom panels: PV diagrams extracted along the horizontal (east--west) and vertical (north--south) directions, respectively. The horizontal red dashed line indicates the adopted systemic velocity of the \hii{} shell, determined empirically from the PV diagrams.}
        \label{fig:13CSpv}
    \end{figure}

    \begin{figure}[ht]
        \centering
        \includegraphics[width=0.5\textwidth]{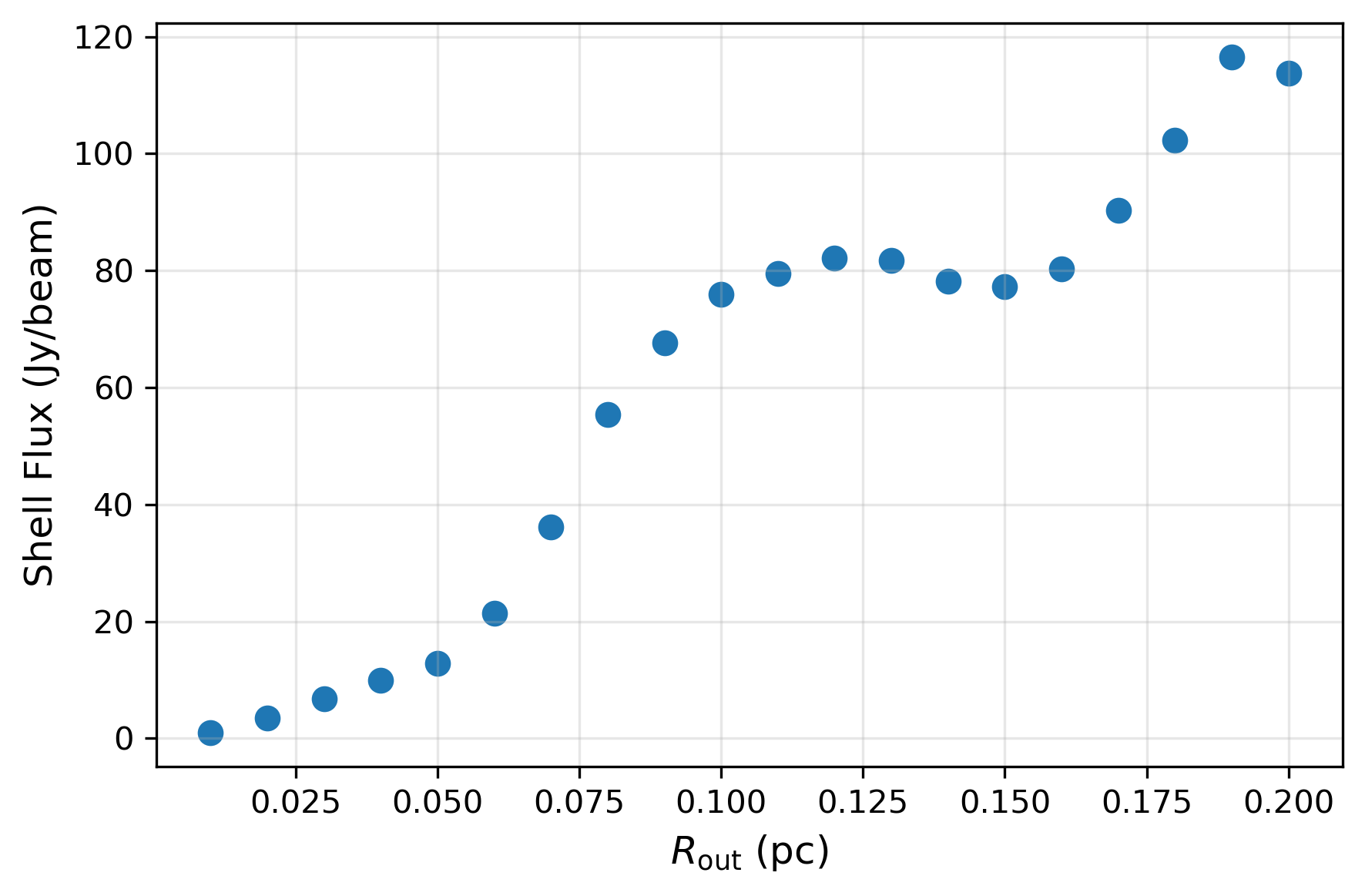}
        \caption{Radial profile of the dust continuum emission around the \hii{} region. The total 1.3~mm flux density within concentric annuli, after subtracting the free--free emission contribution, is plotted as a function of radius. The annuli are centered on the \hii{} region and span radii from 0.01 to 0.20~pc, with a fixed width of 0.01~pc. The flux at each radius represents the integrated emission between the inner and outer boundaries of the corresponding annulus.}
        \label{fig:flux_r}
    \end{figure}

    \begin{figure}[ht]
        \centering
        \includegraphics[width=0.5\textwidth]{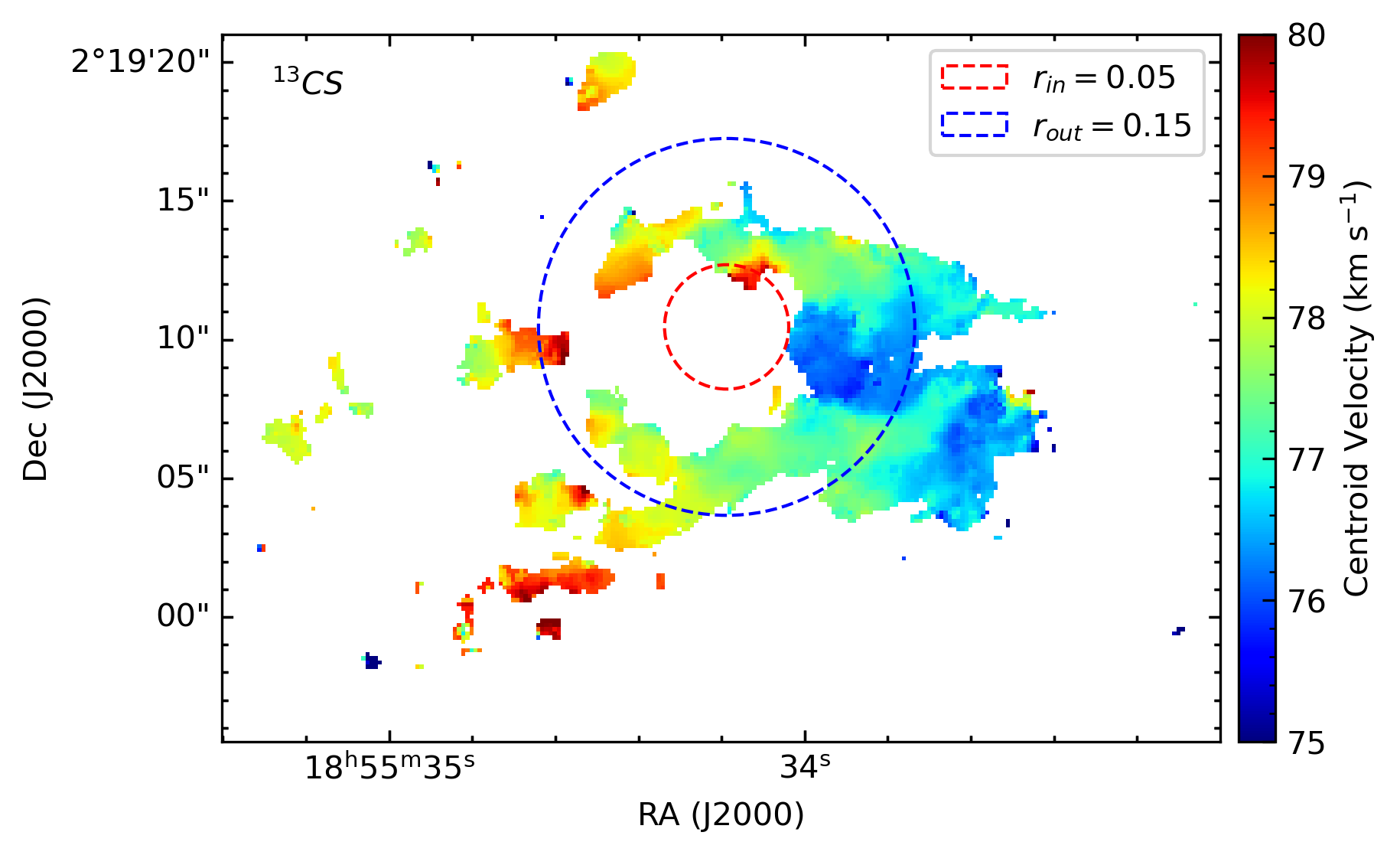}
        \caption{Intensity-weighted mean velocity map of the \tcsff{} emission, integrated over 73--81~km~s$^{-1}$. The red dashed circle marks a region centered on the \hii{} region with a radius of $r_{\rm in}$, while the blue dashed circle indicates a larger region centered on the \hii{} region with a radius of $r_{\rm out}$.}
        \label{fig:13CSshell}
    \end{figure}

    To assess the dynamical state of the \hii{} region in I18530, we estimate both its energy balance and the relevant pressure terms acting on the surrounding gas shell. 
    To probe the dense molecular gas in the vicinity of the \hii{} region, we adopt the \tcsff{} transition, which has a relatively high critical density \rev{and effective excitation density} (see \autoref{tab:tracers}). In our data, the \tcs{} emission delineates a shell-like structure around the ionized region. The center of the \hii{} region was defined from a two-dimensional Gaussian fit to the 1.3~cm continuum emission. We therefore extract PV diagrams along both a horizontal and a vertical cut through the center of the \hii{} region (\autoref{fig:13CSpv}). 
    
    The PV diagrams show different kinematic signatures along the two directions. Along the east--west cut, the emission exhibits a continuous velocity gradient, while the north--south cut displays a characteristic semi-elliptical structure, commonly interpreted as the signature of an expanding shell \citep[e.g.,][]{Arce2011, Butterfield2018}. Although the observed velocity gradient could in principle be attributed to rotation, the presence of a central massive star and an expanding \hii{} region, together with the semi-elliptical PV structure, favors an expansion scenario. The observed east--west velocity gradient may be influenced by the pre-existing large-scale velocity field of the parental cloud or by anisotropic expansion of the shell along the east--west direction. We therefore interpret the observed kinematics primarily as shell expansion.

    Assuming an expanding shell, the expansion velocity is estimated from the maximum velocity offset in the north--south PV diagram, yielding $v_{\rm exp}\approx2.5$~\kms{}. This value should be regarded as a lower limit owing to projection effects.
    
    To estimate the shell radius, we construct a radial profile of the 1.3 mm dust continuum emission centered on the \hii{} region after subtracting the free--free contribution (\autoref{fig:flux_r}). The profile exhibits two distinct breaks at radii of $\sim0.05$ and $\sim0.15$ pc. Between these radii, the annular dust flux increases and then decreases with radius, indicating an enhancement of dust emission in a shell-like structure surrounding the \hii{} region. At larger radii, the profile rises again, likely due to contamination from nearby dense clumps. We therefore adopt $r_{\rm in}=0.05$ pc and $r_{\rm out}=0.15$ pc as the inner and outer radii of the swept-up shell, respectively, as illustrated in \autoref{fig:13CSshell}.

    Assuming a constant expansion velocity, the dynamical age of the shell can be estimated as $t_{\rm dyn} \approx r_{\rm out}/v_{\rm exp}$, which yields $t_{\rm dyn} \approx 5.9 \times 10^4$~yr. These parameters provide the basis for estimating the dynamical properties of the shell, which we analyze below.

        \subsubsection{Energy budget}

        The kinetic energy of the \hii{} region was estimated as
        \begin{equation}
            E_{\rm kin} = \frac{1}{2} M_{\rm shell}\, v_{\rm exp}^2,
        \end{equation}
        where $M_{\rm shell}$ is the mass of the shell, derived from the feathered dust continuum image combining the ALMA 1.3~mm, SMA 1.3~mm, and CSO 1.1~mm data, and $v_{\rm exp}$ denotes the expansion velocity. \rev{The shell gas temperature is derived on a pixel-by-pixel basis from the H$_2$CO line analysis, with a mean value of $\langle T_{\rm gas} \rangle = 56$~K.} The resulting kinetic energy is $E_{\rm kin} = 5.9\times10^{45}~{\rm erg}$. 

        \rev{As shown in \autoref{fig:continuum} (and \autoref{fig:13CSshell}), the weaker 1.3~mm continuum emission toward the eastern boundary may indicate that part of the shell has been blown out, which could lead to an underestimate of the shell mass and kinetic energy. However, the BGPS 1.1~mm continuum data have a much lower angular resolution than the ALMA observations. Therefore, emission from surrounding extended structures is included within the larger beam, which may partially compensate for the missing emission in the ALMA continuum data when estimating the total shell mass.}

        The gravitational potential energy can be approximated as
        \begin{equation}
            E_{\rm grav} = -\,\frac{G\,M_{\rm shell}\,M_{\rm enc}}{R_{\rm eff}},
        \end{equation}
        where $R_{\rm eff}=(r_{\rm in}+r_{\rm out})/2$ is the characteristic radius of the dense gas shell traced by the \tcs{} emission, and $M_{\rm enc}$ is the total mass enclosed within $r_{\rm in}$, including both the stellar mass and the gas mass. Assuming optically thin free–free emission and an electron temperature of 5560~K as estimated by \citet{zhangc2023}, the  stellar mass can be estimated from the ionizing photon rate derived from the 1.3~cm continuum emission.
        The ionizing photon rate is given by \citep{Mezger1974}
        \begin{equation}
            N_{\rm c}' = 4.76 \times 10^{48} \left(\frac{\nu}{\rm GHz}\right)^{0.1} \left(\frac{T_{e}}{\rm K}\right)^{-0.45} \left(\frac{S_\nu}{\rm Jy}\right) \left(\frac{D}{\rm kpc}\right)^2 \ \mathrm{s^{-1}},
        \end{equation}
        where $S_\nu$, D, $\nu$, and $T_e$ are the flux density, distance, observing frequency, and electron temperature, respectively. Using the measured 1.3~cm continuum flux, we obtain $N_{\rm c}'=1.12\times10^{48}$~s$^{-1}$. Assuming that the \hii{} region is powered by a single star, the stellar mass is estimated by comparing the derived ionizing photon rate with published calibrations for ZAMS stars \citep{Davies2011}, yielding a stellar mass of $\sim 20~M_{\odot}$, corresponding to a massive O9 star with a bolometric luminosity of $\sim 10^{4.61}~L_{\odot}$.

        Using this definition, we obtain a gravitational potential energy of $E_{\rm grav} = -1.9 \times 10^{45}$~erg.  
        A comparison between the kinetic and gravitational energies shows that $E_{\rm kin}$ and $|E_{\rm grav}|$ are comparable at the order-of-magnitude level. The total energy $E = E_{\rm grav} + E_{\rm kin} \approx0$ suggests that the system is close to the critical state between bound and unbound.

        The kinetic energy derived for our source ($\sim10^{45}$~erg) is relatively low compared to those reported for other Galactic \hii{} regions, which are typically on the order of $\sim10^{46}$--$10^{49}$~erg \citep[e.g.,][]{Zavagno2007,Peng2010,Bhaswati2024}. This may be related to its early evolutionary stage, as indicated by the shell dynamical age of $\sim6\times10^{4}$~yr.

        \subsubsection{Pressure}

        To further investigate the feedback mechanism of the \hii{} region and its interaction with the surrounding molecular gas, we compare the total internal pressure of the ionized region with the total external pressure exerted by the surrounding molecular shell.

        The internal pressure is assumed to consist of three components: the thermal pressure of the ionized gas ($P_{\rm HII}$), the direct radiation pressure ($P_{\rm dir}$), and the infrared (IR) radiation pressure from dust-reprocessed emission ($P_{\rm IR}$). The external pressure includes the thermal pressure ($P_{\rm ext,th}$), the turbulent pressure ($P_{\rm turb}$), and the gravitational confinement pressure ($P_{\rm grav}$) of the molecular shell. We therefore compare

        \begin{equation}
            P_{\rm int} = P_{\rm HII} + P_{\rm dir} + P_{\rm IR}
            \quad \text{and} \quad
            P_{\rm ext} = P_{\rm ext,th} + P_{\rm turb} + P_{\rm grav}.
        \end{equation}

        The thermal pressure of the swept-up molecular shell is  
        \begin{equation}
        P_{\rm ext,th} = n_{\rm H_2}\, k\, T_{\rm gas},
        \end{equation}
        where $n_{\rm H_2}$ is the molecular hydrogen number density and $k$ is the Boltzmann constant. The molecular hydrogen number density is estimated from the 1.3~mm dust continuum emission, yielding $n_{\rm H_2} = 1.1\times10^{5}$~cm$^{-3}$.
        \rev{Adopting the mean shell temperature of $\langle T_{\rm gas} \rangle = 56$~K,} we obtain $P_{\rm ext,th} = 8.6\times10^{-10}~{\rm dyn~cm^{-2}}$. 

        The turbulent pressure was computed as  
        \begin{equation}
        P_{\rm turb} = \rho\, \sigma_v^{2}.
        \end{equation}
        We adopt the density of the shell $\rho = \mu_{\rm gas}\, m_{p}\, n_{\rm H_2}$, where $\mu_{\rm gas} = 2.37$ is the mean molecular weight. The velocity dispersion $\sigma_v = 0.36~ \kms{}$ is taken from the mean non-thermal velocity dispersion of the ten \tcs{} cores located between $r_{\rm in}$ and $r_{\rm out}$.
        The result is $P_{\rm turb} = 5.8\times10^{-10}~{\rm dyn~cm^{-2}}$.

        The gravitational pressure is estimated as
        \begin{equation}
            P_{\rm grav} = \frac{G M^2}{r_{\rm out} V},
        \end{equation}
        where $V$ is the corresponding volume of the region. 
        The result is $P_{\rm grav} = 1.8\times10^{-8}~{\rm dyn~cm^{-2}}$.

        The thermal pressure inside the ionized gas is given by  
        \begin{equation}
            P_{\rm HII} = 2\, n_e \, k \, T_e.
        \end{equation}
        We estimate the electron density $n_e$ from the 1.3~cm continuum emission within the \hii{} region.
        Assuming spherical symmetry, we measure the integrated flux at 1.3~cm within a radius $r_{in}$ to be $S_\nu = 0.11$~Jy. The electron density is then \citep{Barnes2020}
        \begin{equation}
            n_{\rm e} = 2.576 \times 10^6
            \left( \frac{F_\nu}{\rm Jy} \right)^{0.5}
            \left( \frac{T_e}{\rm K} \right)^{0.175}
            \left( \frac{\nu}{\rm GHz} \right)^{0.05}
            \left( \frac{\theta_{\rm source}}{\rm arcsec} \right)^{-1.5}
            \left( \frac{D}{\rm pc} \right)^{-0.5}
            \ \rm cm^{-3}.
        \end{equation}
        The derived electron density is $n_{\rm e} = 7.1\times10^{3}~{\rm cm^{-3}}$, which is comparable to the value reported by \citet{zhangc2023} ($5.7\times10^{3}~{\rm cm^{-3}}$). Substituting these values, we obtain $P_{\rm HII} = 1.1\times10^{-8}~{\rm dyn~cm^{-2}}$.

        The direct radiation pressure arises from the momentum carried by stellar photons and is given by
        \begin{equation}
            P_{\rm dir} = \frac{3\,L}{4\pi R^{2} c},
        \end{equation}
        where $L$ is the bolometric luminosity of the ionizing source, $R$ is the radius of the \hii{} region (taken as $r_{\rm in}$), and $c$ is the speed of light. This expression assumes that radiation is isotropic and that the radiation field is approximately uniform within the ionized volume. The factor of 3 accounts for the volume-averaged radiation pressure inside a spherical region (e.g., \citealt{Barnes2020}). Substituting the adopted values, we obtain $P_{\rm dir} = 5.2\times10^{-8}~{\rm dyn~cm^{-2}}$.

        In dusty environments, a significant fraction of stellar radiation is absorbed by dust grains and re-emitted in the infrared. This reprocessed radiation can be partially trapped within the shell, contributing an additional pressure component. The infrared radiation pressure is expressed as
        \begin{equation}
            P_{\rm IR} = \frac{u}{3},
        \end{equation}
        where the radiation energy density is written as
        \begin{equation}
            u = U \, u_{\rm ISRF},
        \end{equation}
        where $u_{\rm ISRF} = 8.65\times10^{-13}~{\rm erg~cm^{-3}}$ is the energy density of the local interstellar radiation field, and $U$ is a dimensionless scaling factor describing the intensity of the radiation field relative to the ISRF.

        \begin{figure}[ht]
            \centering
            \includegraphics[width=0.5\textwidth]{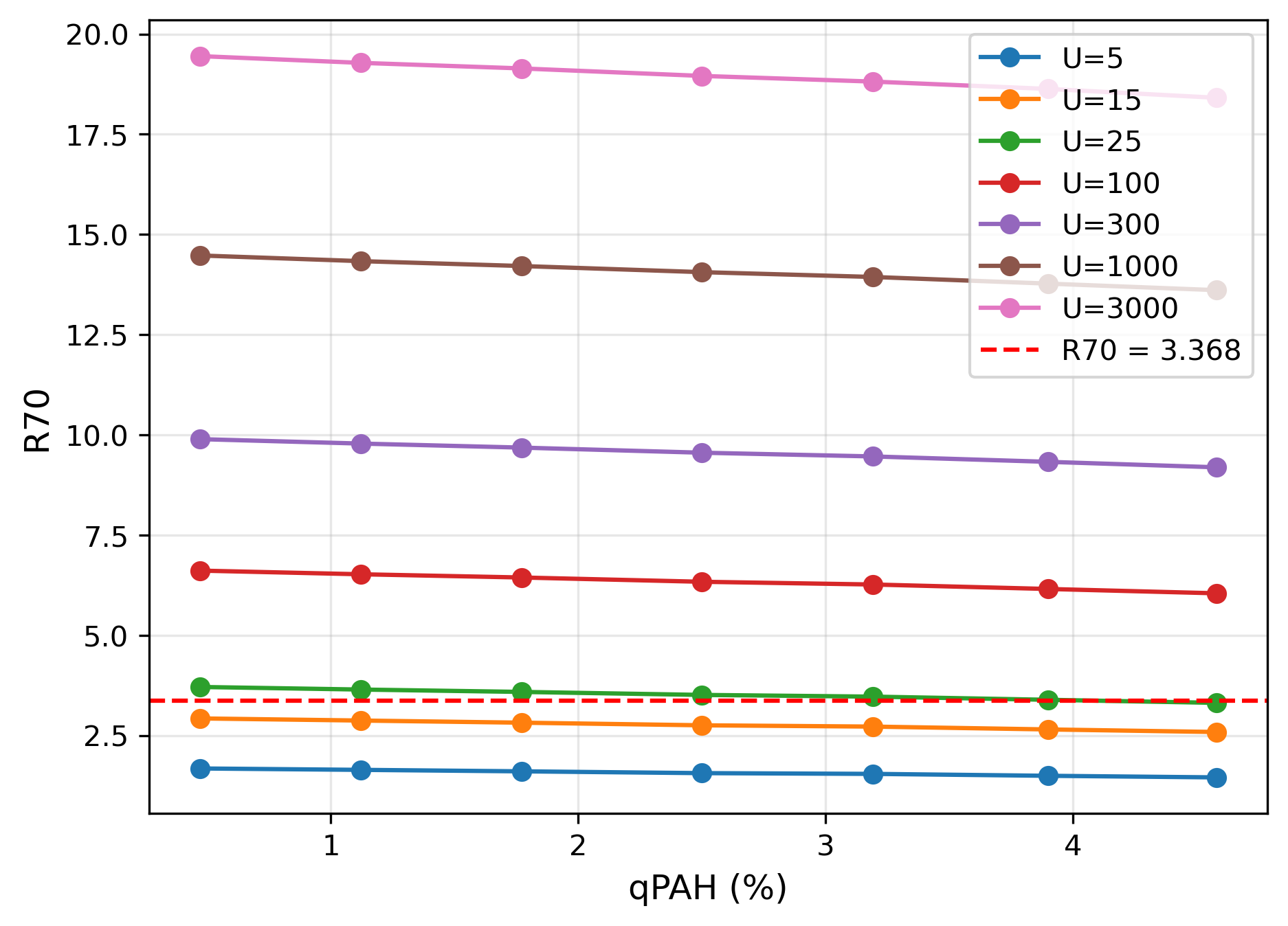}
            \caption{Ratio of $\nu F_{\nu}(70\,\mu{\rm m})$ to $\nu F_{\nu}(160\,\mu{\rm m})$ as a function of $q_{\rm PAH}$ for different radiation field strengths $U$, based on the dust emission models of \citealt{Draine2007}. The red dashed line marks the observed $\nu F_{\nu}(70\,\mu{\rm m}) / \nu F_{\nu}(160\,\mu{\rm m})$ ratio of the \hii{} region.}
            \label{fig:U}
        \end{figure}

        The parameter $U$ is estimated by comparing the observed infrared flux ratio with the dust emission models of \citet{Draine2007}. Specifically, we use the ratio $\nu F_{\nu}(70\,\mu{\rm m}) / \nu F_{\nu}(160\,\mu{\rm m})$ derived from \textit{Herschel} observations within the \hii{} region.

        In the \citet{Draine2007} dust models, the spectral energy distribution (SED) of dust emission is parameterized by the radiation field intensity $U$ and the PAH mass fraction $q_{\rm PAH}$. For a given $q_{\rm PAH}$, the model predicts a monotonic relationship between the ratio $\nu F_{\nu}(70\,\mu{\rm m}) / \nu F_{\nu}(160\,\mu{\rm m})$ and $U$, as warmer dust shifts the emission toward shorter wavelengths, increasing this ratio.

        To determine $U$, we construct a model grid of $\nu F_{\nu}(70\,\mu{\rm m}) / \nu F_{\nu}(160\,\mu{\rm m})$ as a function of $U$ for different values of $q_{\rm PAH}$ (\autoref{fig:U}), following \citet{Draine2007}. The red dashed line in the figure indicates the observed flux ratio measured within the \hii{} region.

        Previous studies have shown that PAH molecules are efficiently destroyed in the harsh ultraviolet radiation fields of \hii{} regions, leading to very low PAH fractions in the majority of Galactic \hii{} regions, typically $q_{\rm PAH} \leq 0.47\%$ \citep{Topchieva2018}. Motivated by this result, we adopt $q_{\rm PAH} = 0.47\%$ as a representative value for our source. At this fixed $q_{\rm PAH}$, the intersection between the observed flux ratio (red dashed line) and the model curve provides a direct estimate of the radiation field intensity. By interpolating along the model grid, we obtain $U = 20.4$. Substituting this value into Equations (19) and (20), we derive an infrared radiation pressure of $P_{\rm IR} = 5.9 \times 10^{-12}\ {\rm dyn\ cm^{-2}}$.
        
        Overall, we find that the outward pressure is dominated by the direct radiation pressure term $P_{\rm dir}$, while the inward pressure is dominated by the gravitational confinement pressure $P_{\rm grav}$. The total external pressure is $P_{\rm ext,th} + P_{\rm turb} + P_{\rm grav} = 1.9 \times 10^{-8}\ {\rm dyn\ cm^{-2}}$, while the total internal pressure is $P_{\rm HII} + P_{\rm dir} + P_{\rm IR} = 6.3 \times 10^{-8}\ {\rm dyn\ cm^{-2}}$. Given the simplified assumptions involved in these estimates, the two pressures are comparable at the order-of-magnitude level, suggesting that the \hii{} region is likely close to pressure equilibrium with the surrounding molecular gas.

        Assuming complete pressure equilibrium between the ionized region and the surrounding molecular gas, we derive a characteristic gas density of $n \sim 3.5 \times 10^{6}\ {\rm cm^{-3}}$. This value is higher than the densities of most DCN and \tcs{} cores in \autoref{tab:coreprop}, implying that a large fraction of the dense cores may not be sufficiently dense to withstand the pressure of the expanding \hii{} region. Consequently, these cores could be displaced and swept outward by the expansion.

        In addition, the ionized gas mass within $r_{\rm out}$ is only $\sim 3.5~M_{\odot}$, compared to $\sim 89~M_{\odot}$ of molecular gas in the surrounding shell. Such a small ionized mass suggests that ongoing photoionization and rapid recombination may have reached a balance in the dense environment, maintaining only a limited amount of gas in the ionized phase at any given time.

        The comparable kinetic and gravitational energies, together with the approximate balance between the outward feedback-driven pressure and the inward confining pressure from the dense gas, suggest that the evolution of the \hii{} region does not follow the picture of classical free expansion. Instead, the ionized region appears to remain strongly regulated by its dense filamentary environment. The relatively low expansion velocity, which is significantly lower than the sound speed of the ionized gas ($\sim10~\kms{}$), further supports a scenario in which the ionized gas is unable to freely expand against the surrounding gravitational potential. Such a scenario is broadly consistent with the gravitationally bound \hii{} region proposed by \citet{Keto2003}, in which the ionized gas is unable to fully overcome the local gravitational potential. At the same time, it also resembles the trapped or choked \hii{} region described by \citet{Matzner2015}, where ionizing feedback is efficiently confined by the high-density surrounding medium and therefore cannot drive large-scale disruption. In this context, the feedback from the \hii{} region in I18530 is likely limited to localized heating and dynamical perturbations on sub-pc scales, while the global filamentary structure can still survive on relatively short evolutionary timescales.

    \subsection{Influence of the \hii{} Region on Surrounding Molecular Cores}\label{subsec:hiiinfluence}
    
\begin{figure*}[ht]
            \centering
            \includegraphics[width=1\textwidth]{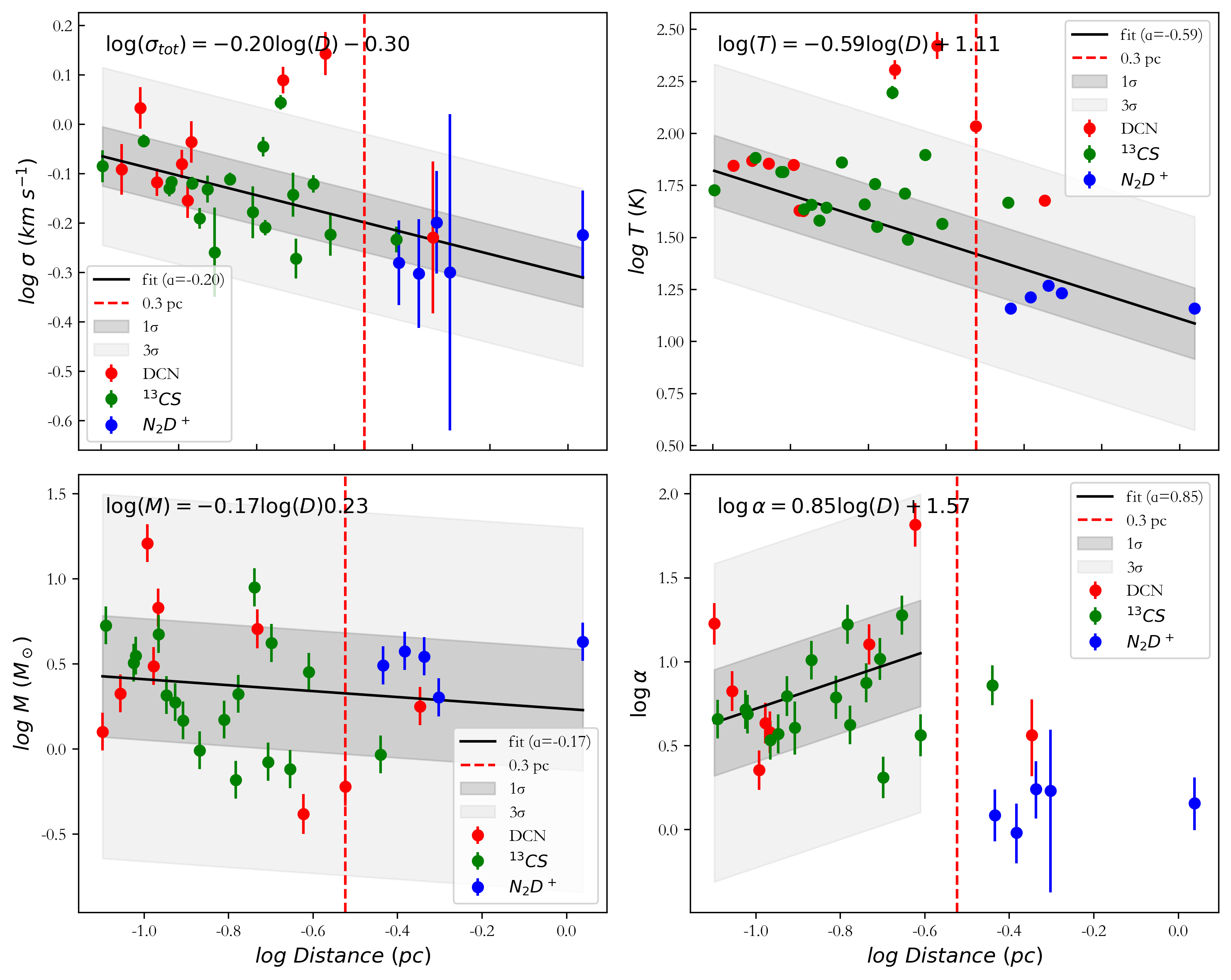}
            \caption{Core velocity dispersion, temperature, mass, and virial parameter as a function of projected distance from the \hii{} region. Red, green, and blue symbols correspond to DCN, \tcs{}, and N$_2$D$^+$ cores, respectively. The red dashed line marks the velocity discontinuity (see \autoref{sec:results}). Linear fits are shown in the corresponding panels. Outliers (see \autoref{subsec:OutlierCores}) are excluded from the velocity dispersion and temperature fits, while the virial parameter is fitted only for cores located near the \hii{} region.}
            \label{fig:velocity dispersion}
    \end{figure*}

    To investigate the impact of the \hii{} region on the surrounding dense cores, we first determine its center by fitting a two-dimensional Gaussian to the 1.3 cm continuum emission. Based on this, we calculate the projected distance of each core from the \hii{} region center. We then examine how the velocity dispersion, temperature, gas mass, and virial parameter vary as a function of the projected distance (\autoref{fig:velocity dispersion}).

    In the absence of external feedback, one would generally expect protostellar cores to exhibit larger line widths than prestellar cores, since internal processes such as gravitational collapse, outflows, and turbulence tend to increase the velocity dispersion. However, in our case, both the velocity dispersion and temperature show a clear decreasing trend with increasing distance from the \hii{} region. Although DCN and \tcs{} cores are generally larger than \ntdp{} cores in spatial scales, which may lead to somewhat broader core-averaged line widths, this effect is unlikely to be dominant. This suggests that feedback from the \hii{} region plays a dominant role in regulating the internal kinematics of the surrounding cores. In particular, ionizing feedback appears to enhance both turbulence and heating in regions close to the \hii{} source \citep[e.g.,][]{Zhang2024}. In contrast, the gas mass shows no significant dependence on projected distance, indicating that the observed trends in kinematic and thermal properties are unlikely to be driven by an underlying primordial mass segregation.
    A possible explanation is that the expansion of the \hii{} region approximately conserves the outward energy flux, such that $4\pi R^2 \rho \sigma^3 \sim {\rm const.}$. Assuming a centrally concentrated density profile of $\rho \propto R^{-1.2}$ \citep{Xu2024raa}, this leads to $\sigma \propto R^{-0.27}$, comparable to the observed trend of approximately $\sigma \propto R^{-0.2}$.

    We note that DCN cores 3 and 6 and \tcs{} core 4, located at a projected distance of $\sim$0.2 pc from the \hii{} region, deviate significantly from the overall trends in both velocity dispersion and temperature. These sources were excluded from the linear fitting and are discussed in detail in \autoref{subsec:OutlierCores}.

    The virial parameter exhibits a two-regime behavior: within $\sim$0.3 pc of the \hii{} region it increases with projected distance, while beyond this scale it drops sharply. Cores at larger distances remain relatively gravitationally bound, suggesting that the dynamical influence of the \hii{} region extends to at least $\sim$0.3 pc.

    \subsection{Outlier Cores}\label{subsec:OutlierCores}

    DCN cores 3 and 6, as well as \tcs{} core 4, show significant deviations from the overall trends in both velocity dispersion and temperature. These cores maintain relatively high velocity dispersions and temperatures even at projected distances where the influence of the \hii{} region is expected to be weak. 

    The temperatures derived from $\mathrm{H_2CO}$ for these cores are systematically higher than those obtained from $\mathrm{NH_3}$ (see \autoref{tab:temperature}) and also deviate from the expected systematic offset reported in \citet{Li2026}. This inconsistency suggests that the temperature structure within these cores is highly inhomogeneous. Moreover, all three cores are associated with $\mathrm{H_2O}$ maser emission, which is widely recognized as a tracer of ongoing star formation activity. Taken together, these observational signatures indicate that the enhanced turbulence and elevated temperatures are most likely dominated by embedded protostellar activity rather than by large-scale feedback from the \hii{} region.

    In terms of spatial distribution, these outlier cores are located near the edge of the dense shell surrounding the \hii{} region. This suggests that their formation may have been influenced by the expansion of the \hii{} region. A coherent scenario can therefore be constructed in which the expansion of the \hii{} region compresses the surrounding molecular gas, leading to the formation of dense cores. Subsequently, embedded protostars evolve within these cores and inject energy locally through feedback processes such as outflows, thereby enhancing the internal velocity dispersion and gas temperature and producing the observed outliers in the global trend.

\section{Conclusions}\label{sec:conc}

    Based on high resolution ALMA Band~6 and VLA K-band observations of the I18530 cloud, we investigate the impact of the \hii{} region on the physical and chemical properties of dense cores along a filamentary structure. Our major findings are as follows:
    \begin{enumerate}

        \item We identify 18, 10, and 5 dense cores from the integrated intensity maps of \tcs{}, DCN, and \ntdp{}, respectively. Most of the identified dense cores have relatively low masses, while only one core (DCN core~7) reaches a mass of $\sim$10~$M_\odot$. This core also exhibits a low virial parameter, suggesting that it is gravitationally bound and has the potential to form a massive star should more gas be accreted from the environment.
        
        \item Dense cores traced by molecular tracers \ntdp{}, DCN, and \tcs{} reveal a clear spatial and physical differentiation along the filament. Cores traced by DCN and \tcs{} are preferentially located close to the \hii{} region and show higher temperatures, velocity dispersions, and virial parameters. \ntdp{} cores are found at larger projected distances and are colder and more quiescent, which may suggest that they are less affected by the feedback from the \hii{} region and are consistent with an earlier evolutionary stage.

        \item The \hii{} region in I18530 has a compact size of $\sim0.1$ pc, an estimated expansion velocity of $\sim2.5~$\kms{}, and a shell dynamical age of $\sim0.06$ Myr, indicating that it is still in a very early evolutionary stage. The outward pressure is dominated by radiation pressure, while the inward confining pressure is primarily governed by the self-gravity of the surrounding dense gas. The comparable inward and outward pressures, together with the expansion velocity being significantly lower than the sound speed of the ionized gas, suggest that the \hii{} region is in a trapped or choked state, in which the ionized feedback is strongly confined by the dense filamentary environment. Assuming pressure equilibrium between the ionized region and the surrounding molecular gas, we estimate a characteristic density of $n \sim 3.5 \times 10^{6}\ {\rm cm^{-3}}$. Gas with densities lower than this value is likely to be swept up or dispersed by the expansion, while denser gas can remain relatively unaffected.
    
        \item In I18530, feedback from massive star formation influences the surrounding gas over a spatial scale of at least $\sim$0.3 pc. Within this region, the feedback manifests primarily as enhanced virial parameters, increased velocity dispersions, and elevated gas temperatures in the dense cores. In contrast, cores located beyond $\sim$0.3 pc remain largely quiescent. The core masses show little dependence on distance from the \hii{} region, providing no evidence that the feedback promotes the formation of more massive dense cores. Notably, several cores located at a projected distance of $\sim$0.2 pc from the \hii{} region exhibit enhanced velocity dispersions and elevated gas temperatures compared to neighboring sources. These cores also show signatures of active star formation, suggesting that local internal energy sources, rather than the \hii{} region alone, contribute to their observed properties.
    \end{enumerate}

\begin{acknowledgements}
We acknowledge helpful discussions with Dr.\ DongDong Zhou (Xinjiang Astronomical Observatory, Chinese Academy of Sciences) on the anomalous hyperfine structure of NH$_3$.
This work has been supported by the National SKA Program of China (2025SKA0140100), the National Natural Science Foundation of China (NSFC) through grant Nos.\ 12273090 and 12322305, the Strategic Priority Research Program of the Chinese Academy of Sciences (CAS) Grant No.\ XDB0800300, and the National Key R\&D Program of China (No.\ 2022YFA1603101).
R.G.M acknowledge support from the UNAM-DGAPA-PAPIIT project IN105225.
This paper makes use of the following ALMA data: ADS/JAO.ALMA\#2017.1.00526.S\@. ALMA is a partnership of ESO (representing its member states), NSF (USA) and NINS (Japan), together with NRC (Canada), MOST and ASIAA (Taiwan), and KASI (Republic of Korea), in cooperation with the Republic of Chile. The Joint ALMA Observatory is operated by ESO, AUI/NRAO and NAOJ\@.
\end{acknowledgements}

\bibliographystyle{aa}
\bibliography{references}

\end{document}